\documentclass[aps,pra,preprint, groupedaddress,amsmath,amssymb,showpacs]{revtex4-2}
\usepackage[english]{babel}
\usepackage{amssymb,bm}
\usepackage{graphicx}
\usepackage{amsmath}
\usepackage{color}
\usepackage{comment}
\usepackage{multirow}
\usepackage[colorlinks=true,citecolor=blue,urlcolor=blue]{hyperref}
\begin{document}
	
\begin{titlepage}

\title{Near-threshold resonances in $e^+e^-$ annihilation}

\author{S.G. Salnikov}
\email{S.G.Salnikov@inp.nsk.su}
\author{A.E. Bondar}
\email{A.E.Bondar@inp.nsk.su}
\author{A.I. Milstein}
\email{A.I.Milstein@inp.nsk.su}
\affiliation{Budker Institute of Nuclear Physics, 630090, Novosibirsk, Russia}

\date{\today}

\begin{abstract}
Near-threshold resonances have been discovered in many hadron pair production processes, resulting in a significant increase in reaction cross sections at low relative velocities of the produced particles. The natural cause of such resonances is the interaction between slow hadrons in the final state. Our review demonstrates that, in virtually all known cases, final-state interactions successfully explain the results of numerous experiments demonstrating a nontrivial energy dependence of cross sections for processes near reaction thresholds. A comprehensive study of near-threshold resonances in various processes can provide new information about the interaction between hadrons at large distances.	
\end{abstract}

\maketitle
\end{titlepage}

\section{Introduction}
Studying the properties of fundamental interactions between elementary
particles is a crucial task in modern physics. In particular,
the experimental and theoretical study
of the details of strong interactions between hadrons at distances about $ 1\,{\text{fm}}$ is of great interest.
One reason for this interest is that precise analytical
calculations of strong interaction effects at low energies  currently pose technical difficulties due to the nonperturbative
nature of quantum chromodynamics. Therefore, in most cases, the description
of interactions between hadrons is based on various phenomenological
approaches. To construct phenomenological models, it is necessary to draw on
experimental information on strong interactions. The
more diverse experimental data used in creating a
model, the greater its predictive power, which, in turn,
can be useful when planning new precision experiments. In the future, when speaking about low energies, we will, as a rule, mean
specifically the region of kinetic energies of produced particles less than 1 GeV.

Various manifestations of the strong interaction between hadrons are studied in a wide variety of processes. For example, the scattering of nucleons (protons and neutrons) from each other, as well as pion-nucleon scattering, are well studied. It is known that at low energies, the quark structure of hadrons does not play a decisive role, and their interaction processes can be described in terms of mesons and baryons (see \cite{Erikson1991Piony}). Recently, much attention has been paid to the study of other processes in which a baryon-antibaryon or meson-antimeson pair is produced in the final state. On the one hand, a wealth of experimental data has appeared on the cross sections of the processes $e^{+}e^{-}\to p\bar{p}$, $e^{+}e^{-}\to n\bar{n}$,
$e^{+}e^{-}\to\Lambda\bar{\Lambda}$, $e^{+}e^{-}\to D\bar{D}$, and
other processes. It has become clear that the energy dependence of the cross sections of such processes
near reaction thresholds is often quite nontrivial.
On the other hand, theorists have proposed various ideas to explain
the observed effects. The dominant idea is the importance of taking into account the interaction between hadrons in the final state. It is the final-state interaction that explains both the sharp increase in cross sections at the reaction threshold,
and the presence of resonance peaks or dips in the energy dependences of the cross sections
near the thresholds.

A proton is undoubtedly the most studied of the hadrons. The structure
and properties of  proton have been studied, in particular, through the scattering
of electrons on protons (beginning with the works~\cite{Hofstadter1955Electron, Bumiller1961Electromagnetic}).
Using Rosenbluth's formula~\cite{Rosenbluth1950High}, it was possible to
derive the electromagnetic form factors of  proton from measurements
of the differential cross section for electron-proton scattering. Knowing the dependence of the form factors
on the magnitude of the momentum transfer during scattering, it is possible, in turn,
to determine the characteristic radius of proton. This method of measuring the proton radius
is still one of the main methods today.

In addition, numerous proton-proton and proton-neutron scattering experiments were performed at various energies and momentum transfers. To theoretically describe the data obtained at low energies, various potential models of nucleon-nucleon interactions were proposed. These include, for example, the Nijmegen potential \cite{nagels1978low,Stoks1994},
Paris potential~\cite{lacombe1980parametrization}, Bonn
potential~\cite{machleidt1987bonn,Machleidt2001}, and potential based
on the chiral quark model~\cite{Entem2000Chiral}.
All such
models are phenomenological. The parameters of the model
potentials are chosen to best describe the available
experimental data for total and differential scattering cross sections.
Then, using these models, it is possible to calculate other quantities
that are difficult to directly observe in experiments, such as scattering phase shifts
or spin-dependent contributions to the scattering cross sections.

With the advent of experimental data on proton-antiproton scattering,
models of nucleon-antinucleon interactions began to develop. An additional
complication, compared to nucleon-nucleon scattering, is that in such processes, the annihilation of initial particles into multiple
meson states is possible. The annihilation cross section is comparable to the elastic scattering cross section due to the large number of available
annihilation channels. For the phenomenological description of annihilation, so-called optical potentials, which contain an imaginary part, are often used. The real
part of nucleon-antinucleon potentials can be obtained by applying the $G$-conjugation transformation to a known nucleon-nucleon potential. However,
the imaginary part of the optical potential can only be determined by
comparison with experimental data on nucleon-antinucleon scattering. The Paris ~\cite{Cote1982,el-bennich2009paris},
Nijmegen \cite{timmermans1994antiproton,zhou2012energy},
 J\"ulich~\cite{hippchen1991meson,mull1991meson,mull1995combined}
nucleon-antinucleon potentials were built in a similar way. Models of nucleon-antinucleon
interaction within the framework of chiral effective field theory~\cite{Kang2014,Dai2017}
and constituent quark models~\cite{Entem2006Interaction,Ortega2025Revisiting} are also being developed.

All of the above-mentioned models provide only a phenomenological description of the available experimental data. Moreover, the nucleon-antinucleon interaction potentials within the different approaches have completely different parametrization. For this reason, different models often describe the experimentally measured properties of the interaction equally well, but they provide significantly different predictions for those characteristics that have not yet been measured experimentally. For example, all nucleon-antinucleon potentials reproduce well the spin-independent contributions to the nucleon-antinucleon scattering cross sections. However, the predictions for the spin-dependent contributions to the nucleon-antinucleon scattering cross sections obtained within the different models differ significantly from each other. \cite{dmitriev2008spin,dmitriev2010spin,Haidenbauer2011Spin} Therefore, it is currently difficult to predict the expected polarization time of an antiproton beam using the filtering method on a polarized target~\cite{csonka1968could}.

The  nucleon-antinucleon interaction is manifested not only in scattering, but also in processes involving the production of nucleon-antinucleon pairs from other initial states. For example, a sharp energy dependence was found for the cross sections of the processes $e^{+}e^{-}\to p\bar{p}$~\cite{Castellano1973,Delcourt1979,Bisello1983,Bisello1990,Aubert2006,Lees2013,Ablikim2015,Akhmetshin2016,Akhmetshin2019,Ablikim2019,Ablikim2020,Ablikim2021b}
and $e^{+}e^{-}\to n\bar{n}$~\cite{Antonelli1998,Achasov2014,Ablikim2021f,Achasov2022,Ablikim2023Measurements}
near the thresholds of these reactions. The most natural explanation for this effect is the nucleon-antinucleon interaction in the final state. Note that the importance of taking into account the interaction between particles in the final state in various processes was already pointed out in the middle of the last century, for example, in Refs.~\cite {Watson1952Effect, Migdal1955Teoriya}. The resonance
form of the dependence of the process probability on the invariant mass of $p\bar{p}$ pair, associated with their interaction, is also observed in the decays of
$\psi$- and $B$-mesons~\cite{Bai2001,Abe2002,Abe2002a,Bai2003,Wang2004b,Aubert2005b,Aubert2006b,Aubert2007a,Ablikim2008,Chen2008,Wei2008,Ablikim2009,Alexander2010,Ablikim2012,Ablikim2013b,Ablikim2013,Ablikim2013a,Ablikim2024Measurement,Ablikim2025Partial}.
Such processes provide another source of information
on the properties of nucleon-antinucleon interactions. While the nucleon form factor in the spacelike momentum transfer region is measured in the nucleon-antinucleon scattering process, in processes involving nucleon-antinucleon pair production, form factors can be measured in the timelike region. Furthermore, as will be shown below, the final-state interaction depends on the wave function of the system at short distances, while the scattering properties are determined by the behavior of the wave function of the system at large distances.

With the development of experimental capabilities, it became clear that the effects of final-state interactions are also observed in processes involving the production of pairs of other particles. We are talking about both baryon-antibaryon pairs ($\Lambda\bar{\Lambda}$,
$\Lambda_{c}\bar{\Lambda}_{c}$ and others~\cite{Aubert2007,Ablikim2018,Ablikim2019c,Ablikim2023,Pakhlova2008,Ablikim2018b,Ablikim2023Measurement,Ablikim2021a,Ablikim2022,Gong2023,Ablikim2024Measurementsa}),
and meson-anti-meson pairs ($D^{(*)}\bar{D}^{(*)}$, $B^{(*)}\bar{B}^{(*)}$
and others
	\cite{Pakhlova2007Measurement, Aubert2007Study, Pakhlova2008Measurement, Cronin-Hennessy2009Measurement, Ablikim2024Precise, Zhukova2018Angular, Ablikim2022Cross, Aubert2009, Mizuk2021, Adachi2024Measurement}). It turned out that such processes also exhibit a nontrivial
dependence of cross sections on energy near reaction thresholds, and often
this dependence has a resonant nature. Moreover, the shape of the observed
resonance peaks is not described by the Breit-Wigner formulas. However,
to explain the experimental results, it is not necessary to assume
the existence of new resonances. The interaction of hadrons in the final state can itself lead to a nontrivial dependence of the cross sections on energy, which explains many effects observed in experiments.

Another interesting effect is the sharp energy dependence
of the cross sections for the production of light mesons near the hadron pair production threshold.
For example, experiments have detected a rapid drop in the cross sections
of the processes $e^{+}e^{-}\to3\left(\pi^{+}\pi^{-}\right)$, $e^{+}e^{-}\to2\left(\pi^{+}\pi^{-}\pi^{0}\right)$ and some others near the nucleon-antinucleon pair production threshold~\cite{Aubert2006a, Akhmetshin2013, Lukin2015, Akhmetshin2019, Aubert2005, Aubert2007c}.
These effects are explained by the production of a virtual nucleon-antinucleon pair near the threshold, followed by annihilation into mesons. The interaction of a nucleon with an antinucleon in the intermediate state leads to a strong energy dependence of the cross sections of these processes.

Thus, a number of near-threshold resonance phenomena observed experimentally are associated with the interaction of hadrons in the intermediate or final state. Therefore, both theoretical and experimental studies of processes in which near-threshold resonances manifest themselves are of great interest. In the monograph \cite{Baz1971Rasseyanie}
it was noted that studying the shape of threshold anomalies in cross sections allows one to obtain a wealth of information about the properties of the system.

It is known from  quantum mechanics (see, for example, ~\cite{landau3rus},
section 132) that in the scattering of slow particles with a large wavelength
compared to the size of the potential, the fine details of the interaction
are unimportant. Specifically, the scattering cross section of slow particles
is expressed in terms of a small number of parameters, such as the scattering length
and the effective interaction radius. This statement is also true
in the description of near-threshold resonances in hadronic systems, since
the resonant dependence of the cross sections of such processes on energy
is often also determined by a small number of parameters. The general laws
that reaction cross sections near the threshold must obey were elucidated
in Refs.~\cite{Wigner1948Behavior, Breit1957Energy, Baz1957Energeticheskaya}.
In addition, in Refs.~\cite {Baz1957Energeticheskaya, Baz1961Energeticheskaya}
the dependence of cross sections on energy below the threshold in the presence of bound states, as well as in the presence of an unstable quasi-stationary
state above the threshold, was considered.

Due to the small number of key parameters needed to describe near-threshold resonances, various phenomenological models can be used. If different models yield similar values for key parameters, then the energy dependences of the process cross sections predicted by these models will be consistent with each other. Therefore, different
groups are developing different approaches to describing the interaction in the final
state. Some authors use interaction potentials in the momentum
representation and find the process amplitudes by solving the Lippmann-Schwinger equation
(see, for example, Refs.~\cite{Haidenbauer2014, Kang2014, Dai2017a, Haidenbauer2021, Yang2024Study, Jia2025Coupledchannel, Ye2025Resonance}).
Other authors use the Watson-Migdal approach~\cite{Haidenbauer2006a},
parameterize the $K$-scattering matrix
	\cite{Husken2022Matrix,Husken2024Polesa}
or use the effective radius expansion for the amplitude~\cite{Zhang2022} and determine the parameter values by comparing the calculated results with experimental data.

Our proposed approach involves parametrization of the interaction potential
between hadrons in a coordinate representation and finding the corresponding
wave function of a hadron pair by solving the Schr\"odinger equation.
The advantage of this approach is that it allows for a clear
interpretation of the resulting interaction potentials, the consideration of
multichannel problems, and the easy inclusion of effects such as
the contribution of the Coulomb potential, violation of isotopic invariance,
and others. The following sections will discuss the application of this
approach to describing the near-threshold behavior of cross sections in various processes.

\section{Production of hadronic pairs near the threshold}

Let's consider a process involving the production of a hadronic pair near the reaction threshold. This could involve a wide variety of processes, including the decay of heavy hadrons with the possible production of other particles in addition to the pair under consideration. For clarity, we will discuss the production of a hadronic pair in electron-positron annihilation: $e^{+}e^{-}\to h\bar{h}$,
where $h$ is a hadron (meson or baryon) having the mass  $M$.
The characteristic mass of hadrons whose production we will consider ranges from one GeV and above. For example, we can talk about nucleons ($M\approx 939\,{\text{MeV}}$),
$\Lambda$-hyperons ($M\approx 1116\,{\text{MeV}}$), $D$-mesons
($M\approx 1870\,{\text{MeV}}$), or $B$-mesons ($M\approx5280\,{\text{MeV}}$).
For a pair of hadrons with a total kinetic energy  $E$ in the center-of-mass frame, the kinematic invariant $s=\left(2M+E\right)^{2}$ (here
and below, we use the natural system of units, $\hbar=c=1$).
Near the hadronic pair production threshold, the conditions $\sqrt{s}\gg E$ are satisfied,
as well as $\sqrt{s}\gg\Lambda_{\text{QCD}}$. Here $\Lambda_{\text{QCD}}\sim300\div350\,{\text{MeV}}$ is the characteristic confinement scale in QCD.

Below, we assume that electron-positron annihilation occurs only through one virtual photon in the intermediate state, since any corrections associated with the possibility of producing more virtual photons are small due to the weakness of the electromagnetic interaction. Then, the process of hadron production in electron-positron annihilation can be considered as follows. First, at small distances of $r\sim1/\sqrt{s}$, a quark and an antiquark are produced from a virtual photon, flying apart in opposite directions. As the quark and antiquark move further apart, the strong interaction between them increases. At distances of $r\sim1/\Lambda_{\text{QCD}}$, the field becomes so strong that new quark-antiquark pairs and gluons begin to be produced from the vacuum. When quarks and antiquarks combine with each other, colorless hadrons are formed, which then fly apart over large distances. The interaction between hadrons continues at large distances, $r\gtrsim1/\Lambda_{\text{QCD}}$, which affects the probability of the formation of such a hadronic system.

Due to the separation of scales ($1/\sqrt{s}\ll1/\Lambda_{\text{QCD}}$),
accounting for final-state interactions in $e^{+}e^{-}\to h\bar{h}$ processes
near the threshold is simplified. The matrix element for the production of $h\bar{h}$ from a
virtual photon can be represented as an integral over the momentum
of hadrons, a product of several factors describing the transition
of a photon to quarks, and then from quarks to interacting hadrons. Since
quark production occurs at short distances, the dependence of the corresponding
momentum factor is determined by the large parameter $\sqrt{s}$,
which changes only slightly with a small change in the kinetic energy
of hadrons. The interaction of hadrons with each other, on the contrary, strongly depends
on their kinetic energy, which changes significantly in the near-threshold
region. Therefore, the integral over hadron momenta will be collected in the region of $p\ll\sqrt{s}$, where the factors describing quark production
can be factored out of the integral as a constant. Thus, the strong dependence of the cross section for the process $e^{+}e^{-}\to h\bar{h}$
on energy near the reaction threshold should be determined, first and foremost, by the interaction of hadrons at distances of $r\gtrsim1/\Lambda_{\text{QCD}}$.

A pair of nonrelativistic hadrons produced in $e^{+}e^{-}$ annihilation can be described using a certain wave function. This wave function can be obtained as a solution to the Sch\"odinger equation in a potential describing the interaction of hadrons with each other at distances of $r\gtrsim1/\Lambda_{\text{QCD}}$.
The explicit form of the equation depends on the hadronic system under consideration, since it is determined by the quantum numbers of the system, the number of possible final states, the forces acting between hadrons, etc.
In the following sections, we will consider both some general examples
of solutions to the Sch\"odinger equation and specific processes involving the production
of a hadron pair near the threshold.

\subsection{Final-state interaction of hadrons}\label{sec:interaction}

To begin, let us consider the influence of final-state interactions
on the amplitude of a process involving the production of a pair of spinless hadrons
in a state with relative orbital angular momentum $L$. The bare amplitude of such a process (i.e., the amplitude calculated without taking into account the interaction between the hadrons) will be proportional to their momentum raised to the power of $L$, and will also have a corresponding dependence on the hadron emission angles. Namely,
\begin{equation}
T_{\text{had}}(\boldsymbol{k})\propto k^{L}\,Y_{Lm}(\hat{\boldsymbol{k}})\,,\label{eq:gen:amplBorn}
\end{equation}
where $\boldsymbol{k}$ is the momentum of hadrons in their center-of-mass frame,
$k=\left|\boldsymbol{k}\right|=\sqrt{ME}$, $Y_{Lm}$ are spherical
functions, and the sign $\hat{\phantom{k}}$
above a vector here and below denotes the corresponding vector of unit
length. For clarity, formula (\ref{eq:gen:amplBorn}) includes only factors that strongly depend on energy and omits other factors
involved in the amplitude of the process $T_{\text{hadron}}$, such as the polarization vectors of particles participating in the reaction, spin factors, momenta
of other particles, etc. Also omitted are the weakly energy-dependent factors describing the production of quarks at small distances.

The  amplitude to detect hadrons with momentum $\boldsymbol{p}$ is given by their wave function in the momentum representation $\Phi^{(-)}_{\boldsymbol{k}}(\boldsymbol{p})$.
This function is obtained by the Fourier transform of the corresponding wave function in the coordinate representation
\begin{equation}
\Phi^{(-)}_{\boldsymbol{k}}(\boldsymbol{p})=\int d^{3}\boldsymbol{r}\,\Psi^{(-)}_{\boldsymbol{k}}(\boldsymbol{r})e^{-i\boldsymbol{pr}}\,.
\end{equation}
Here $\Psi^{(-)}_{\boldsymbol{k}}(\boldsymbol{r})$ is a wave
function of the continuous spectrum, the asymptotic behavior of which at large distances
contains a plane wave and a converging spherical wave, namely
\begin{equation}
\Psi^{(-)}_{\boldsymbol{k}}(\boldsymbol{r})\xrightarrow{r\to\infty}e^{i\boldsymbol{kr}}+f\,\frac{e^{-ikr}}{r}\,,
\end{equation}
where $f$ is some function of  energy and angles. Recall that
it is precisely the function $\Psi^{(-)}_{\boldsymbol{k}}(\boldsymbol{r})$ (or
$\Phi^{(-)}_{\boldsymbol{k}}(\boldsymbol{p})$) that should be used
when calculating the various matrix elements as the wave function
of the final state belonging to the continuous spectrum (see~\cite{landau3rus}, section 136). The hadronic pair production amplitude, taking into account the final-state interaction, $\mathcal{T}_{\text{had}}(\boldsymbol{k})$,  can be obtained from the bare amplitude $T_{\text{had}}(\boldsymbol{p})$ by integrating over all possible values of the hadronic momentum, taking into account
their momentum distribution (see, for example, the review~\cite{Novikov1978Charmonium}
and Refs.~\cite{Dmitriev2006,dmitriev2007final})
\begin{equation}
\mathcal{T}_{\text{had}}(\boldsymbol{k})=\int\frac{d^{3}\boldsymbol{p}}{\left(2\pi\right)^{3}}\,\Phi^{(-)*}_{\boldsymbol{k}}(\boldsymbol{p})\,T_{\text{had}}(\boldsymbol{p})\propto\int\frac{d^{3}\boldsymbol{p}}{\left(2\pi\right)^{3}}\,\Phi^{(-)*}_{\boldsymbol{k}}(\boldsymbol{p})\,p^{L}\,Y_{Lm}(\hat{\boldsymbol{p}})\,.\label{eq:gen:ampl}
\end{equation}

The expansion of the coordinate wave function $\Psi^{(-)*}_{\boldsymbol{k}}(\boldsymbol{r})$
in partial waves can be written as
\begin{equation}
\Psi^{(-)*}_{\boldsymbol{k}}(\boldsymbol{r})=4\pi\sum^{\infty}_{L=0}\sum^{L}_{m=-L}(-i)^{L}\psi^{(R)}_{L}(r)Y_{Lm}(\hat{\boldsymbol{k}})Y^{*}_{Lm}(\hat{\boldsymbol{r}})\,,
\end{equation}
where $\psi^{(R)}_{L}(r)$~ is a regular-at-zero solution of the radial
Sch\"odinger equation in the potential describing the interaction of hadrons.
Here we use the normalization of radial wave functions corresponding to the
asymptotics
\begin{equation}
\psi^{(R)}_{L}(r)\xrightarrow{r\to\infty}\frac{1}{2ikr}\left(S_{L}\,e^{i\left(kr-\pi L/2\right)}-e^{-i\left(kr-\pi L/2\right)}\right),\label{eq:gen:asym}
\end{equation}
where $S_{L}$ are some energy-dependent coefficients. A similar
expansion for the wave function in the momentum representation has the form
\begin{equation}
\Phi^{(-)*}_{\boldsymbol{k}}(\boldsymbol{p})=4\pi\sum^{\infty}_{L=0}\sum^{L}_{m=-L}(-i)^{L}\varphi^{(R)}_{L}(p)Y_{Lm}(\hat{\boldsymbol{k}})Y^{*}_{Lm}(\hat{\boldsymbol{p}})\,.\label{eq:gen:decomp}
\end{equation}
The relationship between radial functions in different representations
is given by the relation
\begin{equation}
\varphi^{(R)}_{L}(p)=4\pi\,i^{L}\int dr\,r^{2}\psi^{(R)}_{L}(r)j_{L}(pr)\,,
\end{equation}
where $j_{L}(x)$~ are the spherical Bessel functions of the first kind. Substituting
the expansion~(\ref{eq:gen:decomp}) into the Eq.(\ref{eq:gen:ampl}),
we obtain (using the properties of Fourier transform and Bessel functions)
\begin{equation}
\mathcal{T}_{\text{had}}(\boldsymbol{k})\propto\frac{(2L+1)!!}{L!}\,Y_{Lm}(\hat{\boldsymbol{k}})\,\frac{\partial^{L}\psi^{(R)}_{L}(0)}{\partial r^{L}}\,,\label{eq:gen:amplInt}
\end{equation}
where $\partial^{L}\psi^{(R)}_{L}(0)/\partial r^{L}$
is the $L$-th derivative of the radial wave function at $r=0$.

Comparing the amplitude~(\ref{eq:gen:amplInt}) with the bare amplitude~(\ref{eq:gen:amplBorn}),
we conclude that taking into account the interaction between hadrons in the final
state leads to a multiplication of the bare amplitude of the process by a factor
\begin{equation}
\mathcal{F}_{L}=\frac{(2L+1)!!}{k^{L}L!}\,\frac{\partial^{L}\psi^{(R)}_{L}(0)}{\partial r^{L}}\,.\label{eq:gen:Fl}
\end{equation}
Then the angularly integrated cross section of the process with the production of a pair
of hadrons will be proportional to
\begin{equation}
\sigma\propto\int d\Omega_{\boldsymbol{k}}\,k\left|\mathcal{T}_{\text{had}}(\boldsymbol{k})\right|^{2}\propto k^{2L+1}\left|\mathcal{F}_{L}\right|^{2},\label{eq:gen:sigel}
\end{equation}
where the factor $k$ in the integral is related to the phase volume of the final
state of the hadronic pair. Here, as before, we do not include factors in the process cross section related to other particles participating
in the reaction, since we are interested in the dependence of the cross section on~$k$.
The factor $k^{2L+1}$ in the cross section~(\ref{eq:gen:sigel}) corresponds
to the energy dependence of the cross section near the reaction threshold, ignoring the interaction
in the final state. The quantity $\left|\mathcal{F}_{L}\right|^{2}$
will henceforth be called the enhancement factor for the cross section for the production
of a hadronic pair in a state with orbital angular momentum~$L$ due to the interaction
in the final state. It can easily be verified that in the absence of interaction
$\mathcal{F}_{L}=1$.

Thus, the amplitude of the process, taking into account the interaction between hadrons
in the final state~(\ref{eq:gen:amplInt}) is proportional
to the wave function
of the hadronic system (or its $L$-th derivative) at zero. This
result appears quite natural, since the wave function
at zero corresponds to the amplitude that the interacting
hadrons could have been at small distances at the moment of production. Similar
expressions were presented, for example, in the review~\cite{Novikov1978Charmonium}
for the amplitudes of charmonium annihilation into various states.
The gain for states with $L=0$ was expressed in terms of the wave
function at zero, (see, for example, Refs.~\cite{Dalkarov1978, Dalkarov2010}).
The change in cross sections due to the Coulomb interaction in the final state is also often described using the Gamow-Sommerfeld-Sakharov factor, which corresponds to the asymptotic behavior of the wave function in the Coulomb field at small distances. Note that the wave function $\psi^{(R)}_{0}(0)$
is inversely proportional to the Jost function, which is also often used to express the cross section enhancement factor due to the interaction in the final state (see, for example, Goldberger-Watson~\cite{GoldbergerWatson}, section 9.3).

To find the wave function of produced pair of spinless hadrons,
we must solve the Sch\"odinger equation in the central potential $V(r)$,
describing their interaction. As is known, the radial wave function
$\psi_{L}(r)$ in a state with orbital angular momentum~$L$ satisfies
the equation
\begin{equation}
\left[\frac{p^{2}_{r}}{M}+V(r)+\frac{L(L+1)}{Mr^{2}}-E\right]\psi_{L}(r)=0\,,\label{eq:gen:Schro1}
\end{equation}
where $p^{2}_{r}=-\frac{1}{r}\frac{\partial^{2}}{\partial r^{2}}r$ is the radial part of the Laplace operator (with the opposite sign). As mentioned above, the solution$\psi^{(R)}_{L}(r)$
should be regular at $r=0$ and have an asymptotic behavior (\ref{eq:gen:asym}) as $r\to\infty$.
In Section \ref{sec:SimpleModel}, we will examine in more detail the model
potential for which equation (\ref{eq:gen:Schro1}) can be solved analytically, and we will find the cross section enhancement coefficient due to the final-state interaction.

\subsection{Interaction of hadrons in an intermediate state}\label{sec:inelastic}

Processes in which the produced pair of real hadrons escapes to infinity
we will call elastic processes. These processes were discussed in the section \ref{sec:interaction},
and the cross section of elastic processes $\sigma_{\text{el}}$ is given by the Eq.~(\ref{eq:gen:sigel}).
Processes of a different kind are also possible, in which a pair of virtual hadrons is first produced in an intermediate state, and then these hadrons annihilate into other particles. We will call such processes
inelastic hadron pair production processes. The interaction
between hadrons in the intermediate state strongly influences the energy dependence
of the cross section of inelastic processes. The total cross section for the production of a hadron pair ($\sigma_{\text{tot}}$) is the sum of the cross section of the elastic process ($\sigma_{\text{el}}$) and the cross section of the inelastic
processes ($\sigma_{\text{inel}}$. In the case of a real
interaction potential $V(r)$, the total cross section coincides with the elastic
cross section above the threshold, and $\sigma_{\text{inel}}=0$ for any
energy except for energies corresponding to possible bound states.
To phenomenological account for the possibility of annihilation of a hadron pair
into other particles, it is convenient to use the so-called optical potential
$V(r)$, which has a negative imaginary part in addition to the real
part. Optical potentials were first proposed in Ref.~\cite{Feshbach1954Model}
to describe the interaction of neutrons with nuclei and have since been applied
in a wide variety of problems. In the presence of an optical potential, the cross section
for inelastic processes will be greater than zero both above and below the threshold for the production of a pair of real hadrons.

The total cross section for the production of a hadronic pair in a state with orbital angular momentum
$L$ can be expressed in terms of the imaginary part of the polarization operator,
which in the nonrelativistic limit is related to the Green's function of the corresponding
Schr\"odinger equation. It can be shown that, up to the factors
which we omitted above when calculating the elastic cross section, the total
cross section is given by the expression
\begin{equation}
\sigma_{\text{tot}}\propto\frac{1}{M}\left(\frac{(2L+1)!!}{L!}\right)^{2}\left.\mbox{Im}{\left[\frac{\partial^{L}}{\partial r^{L}}\frac{\partial^{L}}{\partial r'^{L}}\,\mathcal{D}(r,r'|E)\right]}\right|_{r,r'\to0},\label{eq:gen:sigtot}
\end{equation}
where $\mathcal{D}(r,r'|E)$~ is the Green's function of the radial Schrödinger equation. Namely, it is a solution to the equation
\begin{equation}
\left[\frac{p^{2}_{r}}{M}+V(r)+\frac{L(L+1)}{Mr^{2}}-E\right]\mathcal{D}(r,r'|E)=\frac{1}{rr'}\,\delta(r-r')\,.
\end{equation}
Note that the relationship between the total cross section of a process and the imaginary part of the Green's function was used, for example, in Ref.~\cite{Fadin1987Porogovom},
albeit when solving a somewhat different problem. The Green's function, which has an asymptotic behavior
corresponding to the scattering problem, is expressed in terms of the regular $\psi^{(R)}_{L}$
and irregular $\psi^{(N)}_{L}$ at zero of the solution of the Schr\"odinger equation~(\ref{eq:gen:Schro1}) in the form
\begin{equation}
\mathcal{D}(r,r'|E)=Mk\left[\psi^{(R)}_{L}(r)\psi^{(N)}_{L}(r')\theta(r'-r)+\psi^{(R)}_{L}(r')\psi^{(N)}_{L}(r)\theta(r-r')\right].
\end{equation}
Here $\theta(x)$ is the Heaviside function, and the irregular solution
is normalized according to the asymptotics
\begin{equation}
\psi^{(N)}_{L}(r)\xrightarrow{r\to\infty}\frac{1}{kr}\,e^{i\left(kr-\pi L/2\right)}\,.
\end{equation}
Note that when calculating the Green's function for $E<0$, we must assume that the momentum has a positive imaginary part, that is, $k=i\sqrt{-ME}=i\left|k\right|$.

\subsection{Rectangular potential well as a model of interaction}\label{sec:SimpleModel}
As a simple example of the interaction between hadrons in a state
with $L=0$, we consider a potential in the form of a rectangular well, $V(r)=-V_{0}\cdot\theta(a-r)$,
where $V_{0}$ is the well depth, $a$ is its radius. We will see later that such simple models often describe experimental data well. For such a potential, it is easy to find a solution $\psi^{(R)}_{0}(r)$ of the radial Schr\"odinger equation~(\ref{eq:gen:Schro1})
with $L=0$ that is regular at zero and has an asymptotic behavior~(\ref{eq:gen:asym}) at large distances,
and to obtain the following expression for the factor $\mathcal{F}_{0}$ (see~(\ref{eq:gen:Fl}))
\begin{equation}
\mathcal{F}_{0}=\psi^{(R)}_{0}(0)=\frac{q\,e^{-ika}}{q\cos(qa)-ik\sin(qa)}\,,\qquad k=\sqrt{ME}\,,\qquad q=\sqrt{M(E+V_{0})}\,.\label{eq:gen:psi0}
\end{equation}
The elastic cross section for the production of a hadron pair~(\ref{eq:gen:sigel}) then takes the form
\begin{equation}
\sigma_{\text{el}}\propto k\left|\mathcal{F}_{0}\right|^{2}=k\left|\frac{q}{q\cos(qa)-ik\sin(qa)}\right|^{2}.\label{eq:gen:sigel0}
\end{equation}
Note that in the case of a complex potential, $q$ is also a complex quantity.
For a sufficiently deep potential well ($V_{0}\gg E$,
and therefore $q\gg k$), a resonant enhancement of the cross section will be observed
under the condition
\begin{equation}
q_{0}a=\pi\left(n+\frac{1}{2}\right)+\delta\,,\label{eq:gen:rcond}
\end{equation}
where $q_{0}=\sqrt{MV_{0}}$, $n$ is a non-negative integer,
and~$\left|\delta\right|\ll1$.

If condition~(\ref{eq:gen:rcond}) is satisfied, it is easy to show that
the scattering length of slow particles in the potential under consideration is
$a_{0}=1/q_{0}\delta$ (see, for example,~\cite{landau3rus},  section 133).
In this case, resonant scattering of slow particles will be observed,
since $\left|a_{0}\right|\sim\frac{a}{\delta}\gg a$. A positive
value~$\delta$, for which $a_{0}>0$, corresponds to the presence
of a weakly bound state in the potential well under consideration (shallow
level), and the binding energy is given by the relation $\varepsilon=-1/Ma^{2}_{0}$.
A negative value of $\delta$ (and therefore $a_{0}<0$) corresponds to a so-called virtual level. Recall that the presence of a virtual level
is said to exist when a bound state does not yet exist, but a slight increase in the depth of the potential well leads to the appearance of a shallow level. The energy of a virtual level is, by definition, defined as $\varepsilon=1/Ma^{2}_{0}$.

\begin{figure}\label{fig:flatte}
	\centering
	\includegraphics[totalheight=5.3cm]{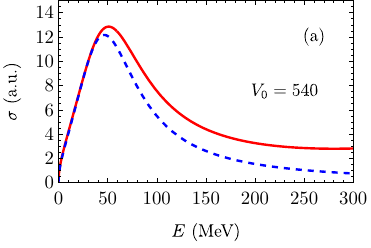}\hfill{}\includegraphics[totalheight=5.3cm]{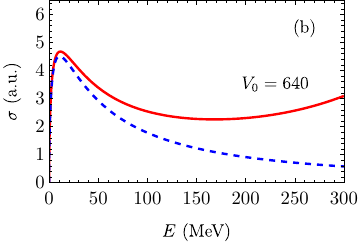}
\caption{Comparison of the energy dependence of the cross sections described by the exact
	formula~(\ref{eq:gen:sigel0}) (solid line) and the Flatt\'e formula~(\ref{eq:gen:flatte})
	(dashed line) for $M=1\,{\text{GeV}}$, $a=2\,{\text{fm}}$ and a few values of  
	  $V_{0}$ (in MeV). Plot (a) corresponds to the value
	$\delta=-0.4$, and plot (b) to the value $\delta=0.25$.}
\end{figure}

\begin{figure}\label{fig:sig0}
	\centering
	\begin{centering}
		\includegraphics[width=1\textwidth]{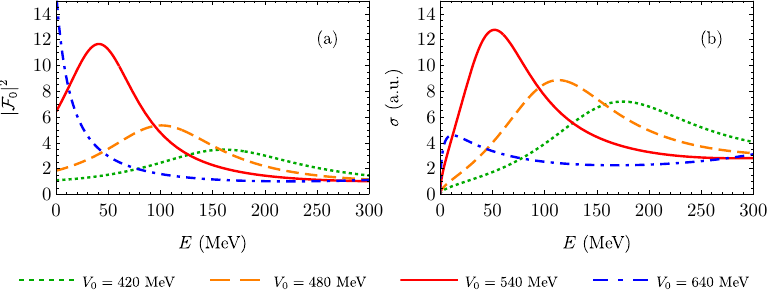}
		\par\end{centering}
	\caption{Dependences of the gain~$\left|\mathcal{F}_{0}\right|^{2}$
		(a) and the cross section for the production of a hadronic pair~(\ref{eq:gen:sigel0}) (b) on
		the energy for $L=0$, $M=1\,{\text{GeV}}$, $a=2\,{\text{fm}}$ and a few
	 values of $V_{0}$ (indicated below the graphs).}
\end{figure}

If the condition~(\ref{eq:gen:rcond}) is satisfied and the potential is real,
the cross section~(\ref{eq:gen:sigel0}) can be reduced to a simpler form
\begin{align}
& \sigma_{\text{el}}\propto\frac{\gamma V_{0}\sqrt{ME}}{\left(E+\varepsilon_{0}\right)^{2}+\gamma E}\,,\nonumber \\
& \varepsilon_{0}=2\kappa V_{0}\delta\,,\qquad\gamma=4\kappa^{2}V_{0}\,,\qquad\kappa=\frac{1}{\pi\left(n+1/2\right)}\,.\label{eq:gen:flatte}
\end{align}
In its structure, this expression corresponds to the Flatt\'e formula ~\cite{Flatte1976}
for the case of a single channel. Note that the energy dependence of the cross section
near the threshold differs significantly from the Breit-Wigner resonance formula.
This is entirely expected, since for the Breit-Wigner formula to be applicable,
the distance to the threshold must be large compared to the resonance width. It is easy to verify that the cross section~(\ref{eq:gen:flatte})
reaches a maximum at an energy of $E=\left|\varepsilon\right|$ for both a shallow level and a virtual level (here
$\varepsilon$ is the energy of the corresponding level). Note also that the exact formula for the cross section~(\ref{eq:gen:sigel0}) has a larger
range of applicability than the Flatt\'e formula~(\ref{eq:gen:flatte}),
since the latter should not be used far from the resonance
peak (see Fig.~\ref{fig:flatte}).
To illustrate the possible influence of final-state interaction
on the energy dependence of the hadronic pair production cross section with $L=0$, we consider some typical examples. Fig.~\ref{fig:sig0}
shows the energy dependence of the gain~$\left|\mathcal{F}_{0}\right|^{2}$
and the hadronic pair production cross section~(\ref{eq:gen:sigel0}) for several values of
 the potential well parameters. For the hadron mass
$M=1\,{\text{GeV}}$ and the potential radius $a=2\,{\text{fm}}$, the third bound state in the potential well appears at a depth
$V_{0}=600\,{\text{MeV}}$. Consequently, at a smaller value
$V_{0}$, a virtual level exists, and a pronounced peak in the gain
is visible. The shallower the well depth, the wider and farther from the threshold this peak becomes. At $V_{0}=640\,{\text{MeV}}$,
conversely, below the threshold there exists a bound state with a binding energy
$\varepsilon=-15{.}4\,{\text{MeV}}$, so the gain
has a sharp maximum right at the threshold.

The cross section for producing a hadronic pair with $L=0$ vanishes at threshold due to the factor $k$ associated with the phase volume of the final state.
Therefore, the cross section maximum is always reached at some positive
energy. As the well depth $V_{0}$ decreases, the virtual level
 is pushed out of the potential well,
resulting in a shift of the cross section peak toward higher energies
(dashed and dotted lines in Fig.~\ref{fig:sig0}). In the presence of
a shallow level below threshold, the cross section peak becomes less pronounced
and is pressed toward the threshold (dash-dotted line in Fig.~\ref{fig:sig0}).
The peak in the hadronic pair production cross section is most pronounced in the case where a virtual level exists near the threshold (solid line in Fig.~\ref{fig:sig0}).

The total cross section for the production of a hadronic pair with $L=0$ (see~(\ref{eq:gen:sigtot}))
can be written as
\begin{equation}
\sigma_{\text{tot}}\propto\mbox{Im}{\left[q\,\frac{q\sin(qa)+ik\cos(qa)}{q\cos(qa)-ik\sin(qa)}\right]}.\label{eq:gen:sigtot0}
\end{equation}
Note that when using an optical interaction potential
containing an imaginary part, the quantity~$q$ becomes
complex. Recall also that the elastic cross section for $L=0$ is given by
Eq.~(\ref{eq:gen:sigel0}) taking into account the complexity of~$q$, and~$\sigma_{\text{inel}}=\sigma_{\text{tot}}-\sigma_{\text{el}}$.

\begin{figure}\label{fig:sigtot0}
	\centering
	\begin{centering}
		\includegraphics[width=1\textwidth]{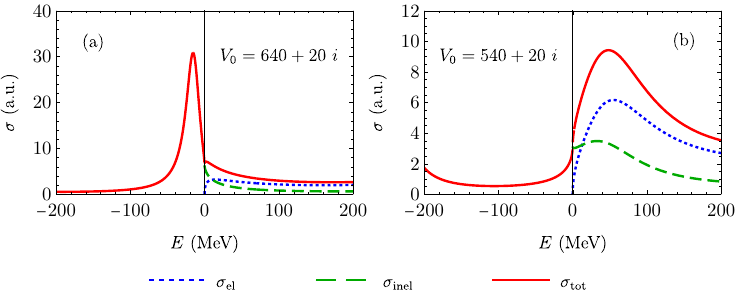}
		\par\end{centering}
	\caption{Energy dependences of the elastic cross section (\ref{eq:gen:sigel0}), total cross section (\ref{eq:gen:sigtot0})
		and inelastic cross section for $L=0$, $M=1\,{\text{GeV}}$,
		$a=2\,{\text{fm}}$ and various values of the optical potential $V_{0}$
		(in MeV). Plot (a) corresponds to a potential containing a shallow level
		below the threshold, while plot (b) corresponds to a potential containing a
		virtual level.}
\end{figure}

Fig. ~\ref{fig:sigtot0} shows the energy dependence of the total, elastic
and inelastic cross sections for different parameters of the optical
potential. Plot (a) corresponds to a potential containing a shallow
level below the threshold, while plot (b) corresponds to a potential containing a virtual level above the threshold. In both cases, the optical
potential $V(r)$ contains a small negative imaginary part to obtain a nonzero cross section for inelastic processes. Because of this,
the bound state below the threshold has a finite, albeit small, width,
which is reflected in the presence of a narrow peak in plot (a) in the total and inelastic
cross sections. The width of the peaks in plot (b), corresponding to the virtual level,
increases due to the addition of the imaginary part to the potential.

\section{The case of multiple reaction channels}

Above, we considered the influence of hadronic interaction on the near-threshold behavior of process cross sections in the case where a hadronic pair can be produced in a single state. However, in real processes, the situation is often more complex. For example, two (or more) thresholds corresponding to the production of pairs of different hadrons are often located quite close in energy (hereinafter, we will discuss different reaction channels). If, in this case, pairs of different hadrons can be produced in states with the same quantum numbers $J^{PC}$, then the interaction between hadrons can lead to transitions between different channels. A typical example of this kind of process is the production of $p\bar{p}$ and $n\bar{n}$ pairs in $e^{+}e^{-}$ annihilation, where the nucleon-antinucleon interaction in the final state
can lead to charge exchange $p\bar{p}\leftrightarrow n\bar{n}$. To theoretically describe such processes near thresholds, it is important to consider all available reaction channels and possible transitions between them.

Consider a certain $n$-channel problem. Suppose that $e^{+}e^{-}$
annihilation can produce pairs of $n$ particles of different types, with
masses $M_{1}$, ..., $M_{n}$, respectively. For definiteness,
assume that $M_{1}<M_{2}<\dots<M_{n}$, and the production thresholds for these
pairs are sufficiently close to each other, that is, 
$\Delta_{i}=2\left(M_{i}-M_{1}\right)\ll M_{i}$. We assume that the quantum numbers of these particles are such that different
pairs of hadrons can be produced in states with relative orbital
momentum~$L$, and transitions between different channels are also possible. The radial wave functions of the $n$-channel problem satisfy a system of $n$ Schr\"odinger equations. It is convenient to write it in matrix form by introducing a column vector of solutions $\Psi(r)=\left(\psi_{1}(r),\,\dots,\,\psi_{n}(r)\right)^{T}$,
where the subscript $T$ denotes transposition. Then the system of radial
Schr\"odinger equations has the form
\begin{equation}
\left[\frac{p^{2}_{r}}{M}+\mathcal{V}(r)+\frac{L(L+1)}{Mr^{2}}-\mathbb{E}\right]\Psi(r)=0\,,\qquad\mathbb{E}_{ij}=\left(E-\Delta_{i}\right)\cdot\delta_{ij}\,,\label{eq:gen:Schro2}
\end{equation}
where the energy $E$ is measured from the threshold for the production of a pair of particles of type 1,
and $M$~is some average mass. The potential matrix $\mathcal{V}(r)$
describes both the interaction of particles of a certain type with each other,
and the transitions between different channels. Linearly independent regular
at zero solutions of Eq.~(\ref{eq:gen:Schro2}), $\Psi^{(R)}_{i}$,
$i=1,\,\dots,\,n$, are determined by the asymptotics of their components with index~$j$
\begin{equation}
\psi^{(R)}_{ij}(r)\xrightarrow{r\to\infty}\frac{1}{2i}\left(S_{ij}\,\chi^{+}_{L}-\delta_{ij}\,\chi^{-}_{L}\right),\qquad\chi^{\pm}_{L}=\frac{1}{k_{j}r}\exp\left[\pm i\left(k_{j}r-\pi L/2\right)\right],
\end{equation}
where $k_{j}=\sqrt{M(E-\Delta_{j})}$, $S_{ij}$~ are some coefficients,
and $\delta_{ij}$~ is the Kronecker delta. Such wave functions describe
states in which particles of type~$j$ are formed at short distances,
and particles of type~$i$ are emitted to infinity.
Then the cross section for the production of a hadron pair of type~$i$ can be written as
\begin{equation}
\sigma^{(i)}\propto k^{2L+1}_{i}\left|\sum^{n}_{j=1}g_{j}\mathcal{F}_{ij}\right|^{2},\qquad\mathcal{F}_{ij}=\frac{(2L+1)!!}{k^{L}_{i}L!}\,\frac{\partial^{L}\psi^{(R)}_{ij}}{\partial r^{L}}(0)\,,\label{eq:gen:sig2}
\end{equation}
where $\mathcal{F}_{ij}$ have the meaning of the final-state interaction gains of the amplitudes associated with the transition
from state~$j$ to state~$i$. The coefficients $g_{j}$ here
are related to the probability amplitudes for the production of the corresponding pairs at
small distances $r\lesssim1/\Lambda_{\text{QCD}}$. In the expression
for the cross sections, as before, we omit some common factors
that do not have a sharp dependence on energy near the reaction thresholds.

To illustrate the characteristic behavior of hadronic pair production cross sections in the case of multiple channels, we will consider some examples. The fundamental patterns of the hadronic pair production cross sections are determined by several factors. First, by the position of the shallow or virtual levels in each channel (in the limit of independent channels). Second, by the amplitudes of the transitions between different channels. Third, by the relationship between the coefficients $g_{j}$ associated with the probabilities of producing different states at small distances.
Below, we will consider the production of hadronic pairs with relative orbital
momentum $L=0$ in the case of two channels. The interaction potential matrix
between hadrons in this case has the form
\begin{equation}
\mathcal{V}(r)=\begin{pmatrix}V_{11} & V_{12}\\
V_{12} & V_{22}
\end{pmatrix}.
\end{equation}
For simplicity, we will use the parametrization of potentials in the form of
rectangular potential wells with depths $U_{ij}$ and a total
radius~$a$, that is
\begin{equation}
V_{ij}(r)=U_{ij}\,\theta(a-r)\,.
\end{equation}

\begin{figure}[!tb]
	\centering
	\begin{centering}
		\includegraphics[totalheight=5.3cm]{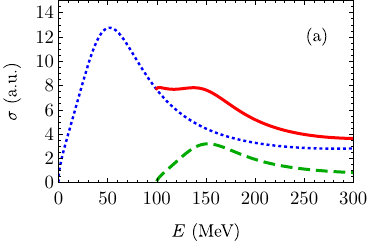}\hspace*{\fill}\includegraphics[totalheight=5.3cm]{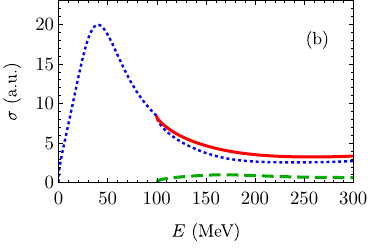}
		\par\end{centering}
	\begin{centering}
		\includegraphics[totalheight=5.3cm]{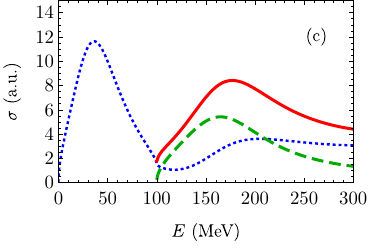}\hspace*{\fill}\includegraphics[totalheight=5.3cm]{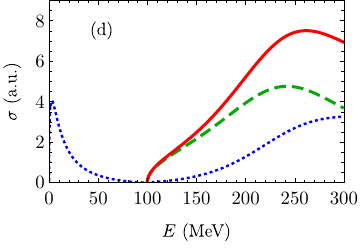}
		\par\end{centering}
	\caption{Energy dependences of the cross sections $\sigma^{(1)}$ (dashed line),
		$\sigma^{(2)}$ (dashed line), and their sum (solid line), see~(\ref{eq:gen:sig2}).
		The model parameters used are $M=1\,{\text{GeV}}$, $\Delta_{2}=100\,{\text{MeV}}$,
		$a=2\,{\text{fm}}$, $U_{11}=U_{22}=-540\,{\text{MeV}}$,
		$g_{1}=1$, $g_{2}=0.5$. In graph (a) $U_{12}=0$, in graph
		(b) $U_{12}=-50\,{\text{MeV}}$, in graph (c) $U_{12}=50\,{\text{MeV}}$,
		and in graph (d) $U_{12}=140\,{\text{MeV}}$.}\label{fig:sig2a}
\end{figure}

First, let's consider the case where both channels contain a virtual
level slightly above the threshold. Using Fig.~\ref{fig:sig0} as a guide,
we choose $U_{11}=U_{22}=-540\,{\text{MeV}}$ for a potential radius
$a=2\,{\text{fm}}$, an average mass $M=1\,{\text{GeV}}$,
and a mass difference of $M_{2}-M_{1}=50\,{\text{MeV}}$. We also fix
the values of the constants $g_{1}=1$ and $g_{2}=0.5$. The energy dependence of the cross sections for the production of hadron pairs of different types, $\sigma^{(1)}$ and $\sigma^{(2)}$, as well as their sums, are shown in Fig.~2a for various values of the off-diagonal potential $U_{12}$. At $U_{12}=0$, there are obviously no transitions between channels, and the total cross section for the production of pairs is simply the sum of the independent cross sections for the production of hadrons of type 1 and type 2, Fig.\ref{fig:sig2a}(a).
 In this case, each of the cross sections $\sigma^{(1)}$ and $\sigma^{(2)}$ contains a peak corresponding to the virtual level. At an off-diagonal potential  $U_{12}=\pm50\,{\text{MeV}}$,
comparable to the difference between the threshold energies, the energy dependence of the cross sections changes significantly. In one case, interference between
channels leads to almost complete suppression of the peak in the second channel,
see Fig.\ref{fig:sig2a}(b). In the other case, conversely, interference
leads to the appearance of a pronounced dip in the total cross section near
the threshold of the second channel, see Fig.\ref{fig:sig2a}(c). Interestingly,
in this case, suppression of the cross section for the production of type 1 hadron pairs is also observed near the second threshold. A further increase in the off-diagonal
potential to $U_{12}=140\,{\text{MeV}}$ leads to an even greater
suppression of the $\sigma^{(1)}$ cross section and a shift of the peak in this cross section closer to the threshold, Fig.\ref{fig:sig2a}(d).

\begin{figure}[!tb]
	\centering
	\begin{centering}
		\includegraphics[totalheight=5.3cm]{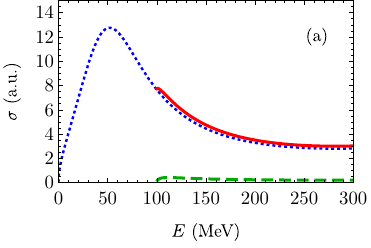}\hspace*{\fill}\includegraphics[totalheight=5.3cm]{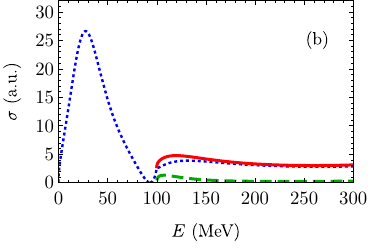}
		\par\end{centering}
	\begin{centering}
		\includegraphics[totalheight=5.3cm]{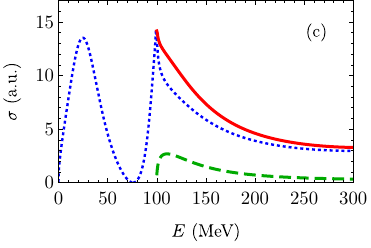}\hspace*{\fill}\includegraphics[totalheight=5.3cm]{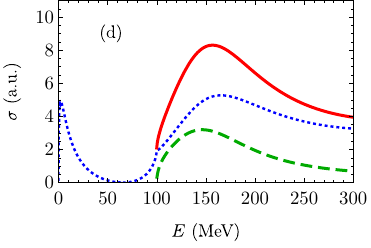}
		\par\end{centering}
	\caption{Energy dependences of the cross sections $\sigma^{(1)}$ (dashed line),
		$\sigma^{(2)}$ (dashed line), and their sum (solid line), see~(\ref{eq:gen:sig2}).
		The model parameters used are $M=1\,{\text{GeV}}$, $\Delta_{2}=100\,{\text{MeV}}$,
		$a=2\,{\text{fm}}$, $U_{11}=-540\,{\text{MeV}}$,
		$U_{22}=-640\,{\text{MeV}}$, $g_{1}=1$, $g_{2}=0.3$.
		In graph (a) $U_{12}=0$, in graph (b) $U_{12}=-50\,{\text{MeV}}$,
		in graph (c) $U_{12}=50\,{\text{MeV}}$, and in graph (d)
		$U_{12}=100\,{\text{MeV}}$.}\label{fig:sig2b}
\end{figure}
Now we choose the potential parameters $U_{11}=-540\,{\text{MeV}}$,
$U_{22}=-640\,{\text{MeV}}$, $a=2\,{\text{fm}}$,
for which there is a virtual level in the first channel and a shallow
level in the second channel. We also fix the values of the constants $g_{1}=1$
and $g_{2}=0.3$. The energy dependence of the cross sections $\sigma^{(1)}$ and
$\sigma^{(2)}$, as well as their sums, are shown in Fig.~\ref{fig:sig2b}
for various values of the off-diagonal potential$U_{12}$. At an off-diagonal potential $U_{12}=0$, the cross section for the production of type 2 particles is practically unnoticeable compared to the cross section for $\sigma^{(1)}$ due to the small value of $g_{2}$, see Fig.\ref{fig:sig2b}(a). However, at $U_{12}=\pm50\,{\text{MeV}}$, the influence of the second channel becomes significant. The manifestation of mixing between the channels is especially noticeable in Fig.\ref{fig:sig2b}(c), where a sharp peak appears in the cross section for the production of type 1 hadrons at the threshold for the production of type 2 hadrons. This effect can be easily explained. Indeed, the bound state of type 2 hadrons existing below the threshold cannot escape to infinity in the absence of transitions between the channels. However,
mixing between hadrons of different types leads to this bound
state decaying into a pair of type 1 hadrons, which already have an energy above their threshold and fly off to infinity. As a result,
a peak appears in the $\sigma^{(1)}$ cross section, located near the second
threshold. A similar picture is observed for $U_{12}=100\,{\text{MeV}}$,
see Fig.~\ref{fig:sig2b}(d).

\section{Process $e^{+}e^{-}\to\Lambda\bar{\Lambda}$}\label{sec:Lambda}
In this and subsequent sections, we discuss some specific processes in which final-state interaction effects manifest themselves.
First, we consider the process of $e^{+}e^{-}$ annihilation into a $\Lambda\bar{\Lambda}$ pair. The possible quantum numbers of the resulting hyperon pair are determined by conservation laws. Since the dominant contribution to the cross section of such processes
comes from the amplitude with one virtual photon in the intermediate state,
the quantum numbers of $\Lambda\bar{\Lambda}$ pair must be $J^{PC}=1^{--}$.
The spatial and charge parities of a system of a baryon and antibaryon
with spins 1/2 are related to the orbital angular momentum $L$ of their relative
motion and their total spin $S$ by the relations $P=\left(-1\right)^{L+1}$,
$C=\left(-1\right)^{L+S}$. From this we obtain that $S=1$, and $L$
can be equal to $0$ or $2$. Near the reaction threshold, the main contribution to the process cross section typically comes from states with the smallest orbital
momentum, i.e., $L=0$. However, as will be shown in Sec. ~\ref{sec:LambdaC},
states with $L=2$ can also make a significant
contribution to the cross section in some cases, even at low energies. Furthermore, mixing of states with $L=0$ and $L=2$ due to tensor forces should lead to
a difference from unity of the ratio of the electric, $G_{E}$, and magnetic,
$G_{M}$, form factors of the baryons.

The cross section of the $e^{+}e^{-}\to\Lambda\bar{\Lambda}$ process was measured
in the range of invariant masses $M_{\Lambda\bar{\Lambda}}$ from the threshold
($2{.}231\,{\text{GeV}}$) to $3{.}08\,{\text{GeV}}$
at the DM2~\cite{Bisello1990}, BaBar~\cite{Aubert2007}
and BESIII~\cite{Ablikim2018,Ablikim2019c,Ablikim2023} detectors. It was found
that the cross section of the process in the immediate vicinity of the threshold is several
times greater than the value that would be expected based on the energy dependence
of the phase volume of the final state. This is a direct indication that for a correct theoretical description of the energy dependence of the cross section of the $e^{+}e^{-}\to\Lambda\bar{\Lambda}$ process, the final-state interaction must be taken into account. In Refs.\cite{Haidenbauer2016,Haidenbauer2021}
the authors considered the interaction between $\Lambda$ and $\bar{\Lambda}$
in the momentum representation, and they were able to obtain a satisfactory
description of the then-available experimental data for the cross section
of the process. Some studies have proposed other approaches to describing the cross section of the $e^{+}e^{-}\to\Lambda\bar{\Lambda}$ process, for example, using a parametrization based on perturbative QCD~\cite{Yang2017},
or using resonance parameters calculated within the quark
model~\cite{Xiao2019}. Here, we consider the application of our approach
to describing the cross section of the $e^{+}e^{-}\to\Lambda\bar{\Lambda}$ process.

The $e^{+}e^{-}\to\Lambda\bar{\Lambda}$ process is the simplest to theoretically describe for several reasons. First, the $\Lambda$-hyperon
consists of $uds$ quarks, with the $ud$ pair existing in a state
with isospin $I=0$. Therefore, the isotopic spin of the $\Lambda$-hyperon
is zero, meaning the $\Lambda$$\bar{\Lambda}$ pair can only exist
in a state with $I=0$. Second, the $\Lambda$-hyperon is
electrically neutral, so the interaction potential between $\Lambda$
and $\bar{\Lambda}$ is determined only by the strong interaction. Third,
currently available experimental data indicate that the ratio of the electromagnetic form factors of the $\Lambda$ hyperon, $\left|G_{E}/G_{M}\right|$,
can be considered equal to unity over a fairly wide energy range
near the threshold. This ratio was measured with good accuracy in the BESIII~\cite{Ablikim2019c} experiment for the invariant mass $M_{\Lambda\bar{\Lambda}}=2{.}396\,{\text{GeV}}$,
and the obtained value, $\left|G_{E}/G_{M}\right|=0{.}96\pm0{.}14$,
is consistent with unity within the experimental uncertainties. The ratio $\left|G_{E}/G_{M}\right|$ was also measured in the BaBar experiment~\cite{Aubert2007}, but the accuracy of these measurements was quite
low. Recently, a new study of the properties of the $\Lambda$ hyperon~\cite{Ablikim2025Unraveling} was conducted at the BESIII detector, which
showed that although the absolute value of the $G_{E}/G_{M}$ ratio is close
to unity, the phase of the form factor ratio still depends on energy.
However, a strong dependence of the phase on energy is observed quite far
from the threshold. Therefore, here we assume that near the threshold, all currently available experimental data for the process $e^{+}e^{-}\to\Lambda\bar{\Lambda}$
can be described without taking into account the contribution of states with $L=2$ (this is confirmed by the results of our calculations).

Thus, the main contribution to the cross section of the $e^{+}e^{-}\to\Lambda\bar{\Lambda}$ process near the threshold should be given by the amplitude for the production of the $\Lambda\bar{\Lambda}$ pair
in a single state with quantum numbers $L=0$, $S=1$, and $I=0$.
Then, the radial wave function $\psi^{(R)}_{0}(r)$ of the $\Lambda\bar{\Lambda}$ pair formed in the $e^{+}e^{-}$ annihilation process satisfies the
Schr\"odinger equation~(\ref{eq:gen:Schro1}) with orbital angular momentum
$L=0$. Since the cross section for the annihilation of the $\Lambda\bar{\Lambda}$ pair into light hadrons is small, the interaction potential $V(r)$ between $\Lambda$
and $\bar{\Lambda}$ can be considered real. As mentioned earlier, the predictions obtained within our approach do not depend strongly on the specific parametrization of the potential. For simplicity of calculations, we
choose a rectangular-well potential,
	$V(r)=-V_{0}\cdot\theta(a-r)$, since this simple parametrization is sufficient for a
good description of the experimental data. The well depth~$V_{0}$
and the potential radius~$a$ are free parameters and should
be determined by comparing the model predictions with the experimental
data.

Taking into account the above, the results of Sec.\ref{sec:SimpleModel} are applicable to the problem of $\Lambda\bar{\Lambda}$ pair production
in $e^{+}e^{-}$ annihilation.
In particular, the amplitude enhancement coefficient due to the final-state interaction, $\mathcal{F}_{0}$, is given by Eq.~(\ref{eq:gen:psi0})
with the $\Lambda$ hyperon mass $M=1115{.}7\,{\text{MeV}}$.
Since the main contribution to the cross section of the $e^{+}e^{-}\to\Lambda\bar{\Lambda}$ process
is determined by the transition through one virtual photon, the amplitude of the process
taking into account the final-state interaction can be written as
\begin{equation}
\mathcal{T}_{\lambda\mu}=\frac{4\pi\alpha}{s}\,g\,F_{D}(s)\,\mathcal{F}_{0}\,\boldsymbol{\epsilon}^{*}_{\lambda}\boldsymbol{e}_{\mu}\,.\label{eq:lam:ampl}
\end{equation}
Here, $\alpha$ is the fine structure constant, $s=\left(2M+E\right)^{2}$,
$E$ is the kinetic energy of the $\Lambda\bar{\Lambda}$ pair, $\boldsymbol{e}_{\mu}$ is the polarization vector of the virtual photon, $\boldsymbol{\epsilon}_{\lambda}$ is the spin wave function of the $\Lambda\bar{\Lambda}$ pair (corresponding
to spin $S=1$), and the indices $\mu$ and $\lambda$ number the possible polarizations.
The factor $g$ is related to the quark production amplitude at small distances
$r\sim1/\sqrt{s}$ and can be considered energy-independent. Its value
is determined by comparison with experimental data. In order to be able to consider a larger energy range, we introduced the dipole form factor of the $\Lambda$ hyperon into the Eq.(\ref{eq:lam:ampl}),
\begin{equation}
F_{D}(s)=\frac{1}{\left(1-s/s_{0}\right)^{2}}\,,\label{eq:lam:FD}
\end{equation}
which takes into account the strong interaction at small distances. This empirical
parametrization of electromagnetic form factors is often used to
describe the interaction of baryons with photons (see, for example,~\cite{Feynman1975Vzaimodeystvie},
 section 22). We fixed the parameter $s_{0}$ equal to $1\,{\text{GeV}}^{2}$.

The differential cross section of the process in the center-of-mass frame is related to the amplitude~(\ref{eq:lam:ampl}) by the relation
\begin{equation}
\left(\frac{d\sigma}{d\Omega}\right)_{\lambda\mu}=\frac{\beta s}{64\pi^{2}}\left|\mathcal{T}_{\lambda\mu}\right|^{2},
\end{equation}
where $\beta=k/M$ is the velocity of $\Lambda$-hyperons. Averaging over the polarizations of a virtual photon and summing over the spin states of the $\Lambda\bar{\Lambda}$ pair can be performed using the relations
\begin{equation}
\frac{1}{2}\sum_{\mu=1,2}e^{i}_{\mu}e^{*j}_{\mu}=\frac{1}{2}\left(\delta^{ij}-n^{i}n^{j}\right),\qquad\sum_{\lambda=1,2,3}\epsilon^{i}_{\lambda}\epsilon^{*j}_{\lambda}=\delta^{ij}\,,
\label{eq:lam:polsum}
\end{equation}
where $\boldsymbol{n}$ is the unit vector directed along the beam collision axis. Note that the summation is only over the two possible polarizations of the virtual photon, since the annihilation of an ultrarelativistic electron and positron occurs only in states
with projections of the total spin onto the collision axis equal to~$\pm1$.
After summation over the spin states of  particles and integration
over the solid angle, we obtain the cross section of the process
\begin{equation}
\sigma=\frac{\pi\beta\alpha^{2}}{s}\,g^{2}F^{2}_{D}(s)\left|\mathcal{F}_{0}\right|^{2},
\end{equation}
where the gain due to the interaction in the final state is given by the Eq.(\ref{eq:gen:psi0}).

\begin{figure}
	\centering
	\includegraphics[totalheight=5.5cm]{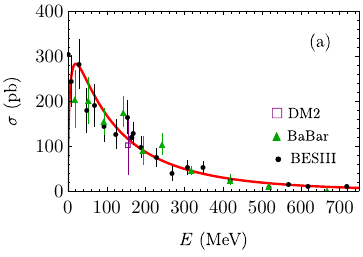}\hfill{}\includegraphics[totalheight=5.5cm]{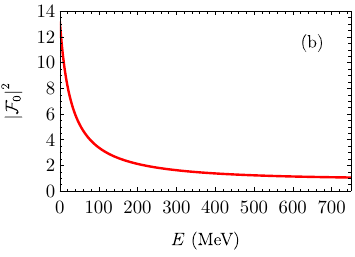}
	
	\caption{Energy dependence of the cross section of the process $e^{+}e^{-}\to\Lambda\bar{\Lambda}$
		(a) and the cross section enhancement factor~(b). Experimental data are
		taken from~\cite{Bisello1990,Aubert2007,Ablikim2018,Ablikim2019c,Ablikim2023}}\label{fig:Lambda}
\end{figure}

To determine the optimal values of the model parameters ($V_{0}$,
$a$, and~$g$), we minimized the value of $\chi^{2}$ (the sum of the squares
of the standard deviations of the theoretical predictions from the experimental values
of the cross section). We used data obtained at the DM2~\cite{Bisello1990},
BaBar~\cite{Aubert2007}, and BESIII~\cite{Ablikim2018,Ablikim2019c,Ablikim2023} detectors. As a result, the following parameter values were found: $V_{0}=584\,{\text{MeV}}$,
$a=0{.}45\,{\text{fm}}$, $g=2{}37$ (these results were
presented in our paper~\cite{Milshteyn2023Estestvennoe}). The corresponding
value is $\chi^{2}/N_{\mathrm{df}}=37{}6/29=1{.}3$, where $N_{\mathrm{df}}$ is the number of degrees of freedom, equal to the difference between the number of experimental
points and the number of free parameters of the model. A comparison of our results
for the cross section of the $e^{+}e^{-}\to\Lambda\bar{\Lambda}$ process with experimental
data is shown in Fig.\ref{fig:Lambda}(a). Fig.\ref{fig:Lambda}(b)
shows the energy dependence of the cross section enhancement coefficient due to the final-state interaction. Clearly, accounting for the interaction
between $\Lambda$ and $\bar{\Lambda}$ is necessary for a correct
description of the experimental data.

Our model predicts the existence of a bound state of $\Lambda$ and $\bar{\Lambda}$ with a binding energy of $E_{0}\approx-30\,{\text{MeV}}$.
Such a state, if it truly exists, could manifest itself in other processes, leading to a sharp dependence of the cross sections on
energy. Note that in the experimental data~\cite{Aubert2005,Aubert2007c,Chang2009,Ablikim2019b,Ablikim2019d,Ablikim2020a,Ablikim2020b,Ablikim2021g,Xia2022,Ablikim2023Measurementa}
there are some indications of anomalous behavior of the cross sections of the processes
$e^{+}e^{-}\to K^{+}K^{-}\pi^{+}\pi^{-}$, $e^{+}e^{-}\to2\left(K^{+}K^{-}\right)$,
$e^{+}e^{-}\to\phi K^{+}K^{-}$ and others at $\sqrt{s}=2.2\,{GeV}$,
which exactly corresponds to the energy $E\approx-30\,{\text{MeV}}$.
The analysis of $p\bar{\Lambda}$ and $\Lambda\bar{\Lambda}$ interactions,
conducted in Ref.~\cite{Sarti2025Novel}, also qualitatively corresponds
to the assumption of the existence of a bound state~$\Lambda\bar{\Lambda}$.
The assumption of the possible existence of a bound state in the system
$\Lambda\bar{\Lambda}$ was also put forward in the Ref.\cite{Carbonell1993Possible}
based on an analysis of data for the cross section of the process $p\bar{p}\to\Lambda\bar{\Lambda}$.

\section{Process $e^{+}e^{-}\to\Lambda_{c}\bar{\Lambda}_{c}$}\label{sec:LambdaC}

In the $e^{+}e^{-}\to\Lambda_{c}\bar{\Lambda}_{c}$ process, effects related to the final-state interaction are also observed. The experiments
Belle~\cite{Pakhlova2008} and BESIII~\cite{Ablikim2018b,Ablikim2023Measurement}
measured the energy dependence of the cross section for this process, as well as the ratio of the electromagnetic form factors of the $\Lambda_{c}$ baryon. There have also been theoretical studies devoted to describing these experimental
data using different approaches. Thus, the authors of Ref.\cite{Amoroso2021}
attempted to reproduce the energy dependence of the cross section observed in experiments by parametrization of the resonant contribution using a Gaussian distribution. In Refs. ~\cite{Dai2017a,Guo2024Study} the Lippmann-Schwinger equation was used to study the interaction between $\Lambda_{c}$ and $\bar{\Lambda}_{c}$ in the momentum representation, while in Refs.\cite{Chen2024Lambda,Chen2025Electromagnetic} the electromagnetic form factors of $\Lambda_{c}$ baryon were presented  using Breit-Wigner distributions.

The theoretical description of the process $e^{+}e^{-}\to\Lambda_{c}\bar{\Lambda}_{c}$ within our approach has much in common with the description of the process $e^{+}e^{-}\to\Lambda\bar{\Lambda}$ discussed in the previous section, but there are also important differences. The first
obvious difference is due to the fact that $\Lambda_{c}$ has a charge $+1$,
meaning that the Coulomb interaction between $\Lambda_{c}$
and~$\bar{\Lambda}_{c}$ must be taken into account. The second difference stems from the fact that in the
BESIII experiment~\cite{Ablikim2018b,Ablikim2023Measurement}
a significant deviation from unity was obtained for the ratio of the electromagnetic
form factors~$\Lambda_{c}$ near the threshold. This suggests the need to take into account not only the amplitude for the production of states $\Lambda_{c}\bar{\Lambda}_{c}$
with orbital angular momentum $L=0$, but also the contribution of states with $L=2$.

As is known from nuclear physics (see, for example,\cite{Erikson1991Piony}),
the interaction potential between baryons contains a contribution from so-called
tensor forces, $V_{T}(\boldsymbol{r})=V_{T}(r)S_{12}$. The tensor
operator $S_{12}$ is defined by the relations
\begin{equation}
S_{12}=3\left(\boldsymbol{\sigma}_{1}\boldsymbol{n}\right)\left(\boldsymbol{\sigma}_{2}\boldsymbol{n}\right)-\left(\boldsymbol{\sigma}_{1}\boldsymbol{\sigma}_{2}\right)=2\left[3\left(\boldsymbol{S}\boldsymbol{n}\right)^{2}-\boldsymbol{S}^{2}\right],
\end{equation}
where $\boldsymbol{n}=\boldsymbol{r}/r$ is the unit vector, $\boldsymbol{\sigma}_{1}$ and
$\boldsymbol{\sigma}_{2}$ are the Pauli matrices (doubled spin operators) for the first and second particles, and $\boldsymbol{S}=\frac{1}{2}\left(\boldsymbol{\sigma}_{1}+\boldsymbol{\sigma}_{2}\right)$ is the total spin operator. The tensor operator preserves the value of the total
spin of the system~$S$ and the total angular momentum~$J$, but can change the value of the
orbital angular momentum~$L$ by an amount~$\pm2$. Thus, the states
of a system with spin $S=1$, total angular momentum~$J=1$, and orbital angular momenta
$L=0$ and $L=2$ are mixed due to the action of tensor forces.

To describe the interaction between baryons with account for the tensor forces, we need to consider a system of Schr\"odinger equations for two coupled partial waves. It can be written as
\begin{equation}
\left[\frac{p^{2}_{r}}{M}+\mathcal{V}(r)-E\right]\Psi_{\nu}(r)=0\,,\label{eq:lamc:Schro}
\end{equation}
where $\Psi_{\nu}(r)=\left(u_{\nu}(r),w_{\nu}(r)\right)^{T}$ is a two-component column vector of solutions to the equation, and the index $\nu$, equal to 1 or 2, numbers linearly independent solutions. The upper component,
$u(r)$, corresponds to the orbital angular momentum $L=0$, and the lower component,
$w(r)$, corresponds to $L=2$. The mass $M=2286{.}5\,{\text{MeV}}$ in
this section denotes the mass of the $\Lambda_{c}$ baryon, and $E$ is the kinetic energy of the $\Lambda_{c}\bar{\Lambda}_{c}$ pair. The interaction potential (taking into account Coulomb forces and the centrifugal contribution)
can be represented as a matrix
\begin{equation}
\mathcal{V}(r)=\begin{pmatrix}-\frac{\alpha}{r}+V_{S} & 2\sqrt{2}\,V_{T}\\
2\sqrt{2}\,V_{T} & -\frac{\alpha}{r}+\frac{6}{Mr^{2}}+V_{D}-2V_{T}
\end{pmatrix},\label{eq:lamc:pot}
\end{equation}
where $V_{S}$ is the interaction potential in the state with $L=0$,
$V_{D}$ is the interaction potential in the state with $L=2$, and~$V_{T}$~ is the tensor potential. Linearly independent solutions of the system, $\Psi_{\nu}(r)$,
are determined by their asymptotics at infinity,
\begin{align}
& u_{1}(r)\xrightarrow{r\to\infty}\frac{1}{2i}\left(S_{11}\,\chi^{+}_{0}-\chi^{-}_{0}\right), &  & w_{1}(r)\xrightarrow{r\to\infty}\frac{1}{2i}\,S_{12}\,\chi^{+}_{2}\,,\nonumber \\
& u_{2}(r)\xrightarrow{r\to\infty}\frac{1}{2i}\,S_{21}\,\chi^{+}_{0}\,, &  & w_{2}(r)\xrightarrow{r\to\infty}\frac{1}{2i}\left(S_{22}\,\chi^{+}_{2}-\chi^{-}_{2}\right),
\end{align}
where $S_{ij}$~ are some coefficients. In this case, in the presence of a
Coulomb potential, the diverging and converging waves contain an additional
Coulomb phase, namely
\begin{align}
& \chi^{\pm}_{l}=\frac{1}{kr}\exp\left[\pm i\left(kr-l\pi/2+\eta\ln\left(2kr\right)+\sigma_{l}\right)\right],\nonumber \\
& \sigma_{l}=\frac{i}{2}\ln\frac{\Gamma(1+l+i\eta)}{\Gamma(1+l-i\eta)}\,,\qquad\eta=\frac{M\alpha}{2k}\,.
\end{align}

The amplitude for the production of a pair of $\Lambda_{c}$-baryons in $e^{+}e^{-}$~annihilation,
taking into account the final-state interaction, can be written
as (see, for example,~\cite{Dmitriev2014Isoscalara})
\begin{equation}
\mathcal{T}_{\lambda\mu}(\boldsymbol{k})=\frac{4\pi\alpha}{s}\,G_{S}F_{D}(s)\left\{ u_{1}(0)\bigl(\boldsymbol{\epsilon}^{*}_{\lambda}\boldsymbol{e}_{\mu}\bigr)-\frac{u_{2}(0)}{\sqrt{2}}\left[\bigl(\boldsymbol{\epsilon}^{*}_{\lambda}\boldsymbol{e}_{\mu}\bigr)-3\bigl(\hat{\boldsymbol{k}}\boldsymbol{\epsilon}^{*}_{\lambda}\bigr)\bigl(\hat{\boldsymbol{k}}\boldsymbol{e}_{\mu}\bigr)\right]\right\} ,
\end{equation}
where $\boldsymbol{e}_{\mu}$ is the polarization vector of the virtual
photon, and $\boldsymbol{\epsilon}_{\lambda}$ is the spin wave
function of the pair $\Lambda_{c}\bar{\Lambda}_{c}$ (with total spin $S=1$).
The constant $G_{S}$ corresponds to the hadron production amplitude at small
distances with $L=0$, and the dipole form factor $F_{D}(s)$ is defined
according to Eq.~(\ref{eq:lam:FD}) with the same value of the parameter
$s_{0}=1\,{\text{GeV}}^{2}$. The non-zero value of $u_{2}(0)$
arises due to the influence of tensor forces, which results in a noticeable
contribution of the $D$-wave to the amplitude of the process even in an approximation where
at small distances the $S$-wave is predominantly present.

The differential cross section for the production of a pair $\Lambda_{c}\bar{\Lambda}_{c}$
in $e^{+}e^{-}$~annihilation, taking into account the rules of summation over polarizations~(\ref{eq:lam:polsum})
can be reduced to the form
\begin{equation}
\frac{d\sigma}{d\Omega}=\frac{\beta s}{128\pi^{2}}\sum_{\lambda\mu}\left|\mathcal{T}_{\lambda\mu}\right|^{2}=\frac{\beta\alpha^{2}}{4s}\left[\left|G_{M}\right|^{2}\left(1+\cos^{2}\theta\right)+\frac{4M^{2}}{s}\left|G_{E}\right|^{2}\sin^{2}\theta\right].
\end{equation}
Here $\theta$~ is the angle between the axis of the electron-positron collision
and the direction of emission of one of the baryons, and $G_{E}$ and $G_{M}$~ are the electric and magnetic form factors of the $\Lambda_{c}$-baryon, determined by
the relations
\begin{align}
& \frac{2M}{\sqrt{s}}\,G_{E}=\frac{G_{S}F_{D}(s)}{\sqrt{2}}\left(u_{1}(0)+\sqrt{2}\,u_{2}(0)\right),\nonumber \\
& G_{M}=\frac{G_{S}F_{D}(s)}{\sqrt{2}}\left(u_{1}(0)-\frac{1}{\sqrt{2}}\,u_{2}(0)\right).\label{eq:lamc:GeGm}
\end{align}
The cross section for the production of the pair $\Lambda_{c}\bar{\Lambda}_{c}$ integrated over angles is
\begin{equation}
\sigma=\frac{2\pi\beta\alpha^{2}}{3s}\left(2\left|G_{M}\right|^{2}+\frac{4M^{2}}{s}\left|G_{E}\right|^{2}\right)=\frac{\pi\beta\alpha^{2}}{s}\,G^{2}_{S}F^{2}_{D}(s)\left(\left|u_{1}(0)\right|^{2}+\left|u_{2}(0)\right|^{2}\right).\label{eq:lamc:sig}
\end{equation}
It is also evident from Eq.~(\ref{eq:lamc:GeGm}) that in the near-threshold
energy region (that is, $E\ll2M$) the ratio of electromagnetic form factors
has the form
\begin{equation}
\frac{G_{E}}{G_{M}}=\frac{u_{1}(0)+\sqrt{2}\,u_{2}(0)}{u_{1}(0)-\frac{1}{\sqrt{2}}\,u_{2}(0)}\,.
\end{equation}
This ratio is determined solely by the properties of the interaction in the final state and does not contain any other normalization factors. It follows from this that the deviation of the $G_{E}/G_{M}$ ratio from unity near the threshold arises solely due to the influence of tensor forces, which lead to a mixing of states with orbital angular momenta $L=0$ and $L=2$.

Taking into account the above, to describe the interaction of $\Lambda_{c}$
and $\bar{\Lambda}_{c}$ in the final state, we need to find the potentials
$V_{S}(r)$, $V_{D}(r)$, and $V_{T}(r)$ (see~(\ref{eq:lamc:pot})).
Since at low kinetic energies, particles are not particularly sensitive
to the details of the interaction, we use a simple model potential
 for the $\Lambda_{c}\bar{\Lambda}_{c}$ interaction (the corresponding results
were presented in our paper~\cite{Salnikov2023b}). All three
potentials ($S$-wave, $D$-wave, and tensor) were parameterized
as rectangular wells
\begin{equation}
V_{i}(r)=U_{i}\cdot\theta(a_{i}-r)\,,\qquad i=S,\,D,\,T\,.\label{eq:lamc:pots}
\end{equation}
In addition, for convenience of calculations, the tensor potential was regularized
at small distances by the factor
\begin{equation}
F(r)=\frac{\left(br\right)^{2}}{1+\left(br\right)^{2}}\,,
\end{equation}
where $b=10\,{\text{fm}^{-1}}$ is chosen. This regularization
ensures consistency of the asymptotics of the solutions
$u(r)$ and $w(r)$, Eq.(\ref{eq:lamc:Schro}), at small distances (namely, $u(r)\sim\textrm{const}$,
$w(r)\sim r^{2}$). Note that our results are practically independent
of the value of the parameter~$b$.

To determine the interaction potential parameters, we used experimental data obtained with the BESIII detector~\cite{Ablikim2018b,Ablikim2023Measurement}.
Note that the cross section for the $e^{+}e^{-}\to\Lambda_{c}\bar{\Lambda}_{c}$ process
was also measured with the Belle detector~\cite{Pakhlova2008}; however,
data from these two detectors contradict each other in the range
of invariant masses from $4{.}6\,{\text{GeV}}$ to $4{.}65\,{\text{GeV}}$.
Furthermore, the earlier Belle data have significantly larger
experimental uncertainties than the later BESIII data. Therefore, we decided to limit our consideration to the experimental data obtained at BESIII. The optimal potential parameters (\ref{eq:lamc:pots})
were determined by minimizing the $\chi^{2}$ calculated from the experimental
data for the cross section of the process $e^{+}e^{-}\to\Lambda_{c}\bar{\Lambda}_{c}$,
as well as the ratio $\left|G_{E}/G_{M}\right|$. The free parameters
of our model include the radii and depths of the potential wells, as well as the bare $S$-wave form factor $G_{S}$,
which determines the overall normalization of the cross section (\ref{eq:lamc:sig}). The interaction potentials can be considered real, since the annihilation cross section of $\Lambda_{c}\bar{\Lambda}_{c}$ into light mesons is small. The obtained values
of the parameters are listed in Table~\ref{tab:LL}, and the corresponding value
$\chi^{2}/N_{\mathrm{df}}=46{.}9/34=1{.}38$.

\tabcolsep=1em
\begin{table}
	\centering
	\caption{Parameters of the model describing the interaction of $\Lambda_{c}$ and $\bar{\Lambda}_{c}$.}\label{tab:LL}

	\centering{}
	\begin{tabular}{|l|c|c|c|}
		\hline
		& $V_{S}$ & $V_{D}$ & $V_{T}$\tabularnewline
		\hline
		\hline
		$U\,(\text{MeV})$ & $-1180$ & $-170$ & $-64$\tabularnewline
		\hline
		$a\,{(\text{fm})}$ & $0{,}98$ & $1{,}93$ & $0{,}75$\tabularnewline
		\hline
		$G_{S}$ & \multicolumn{3}{c|}{$155{,}4$}\tabularnewline
		\hline
	\end{tabular}
\end{table}

\begin{figure}[!tb]
	\centering
	\begin{centering}
		\includegraphics[totalheight=5.2cm]{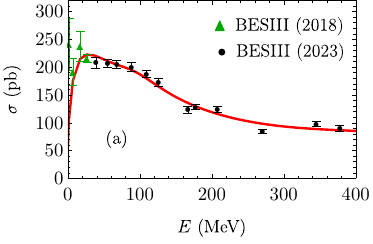}\hfill{}\includegraphics[totalheight=5.2cm]{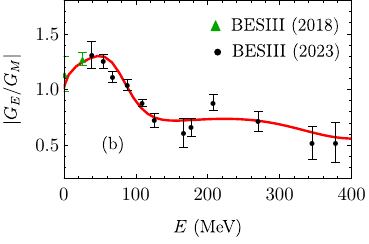}
		\par\end{centering}
	\begin{centering}
		\includegraphics[totalheight=5.2cm]{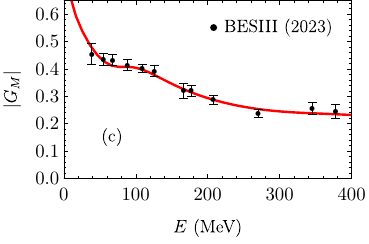}\hfill{}\includegraphics[totalheight=5.2cm]{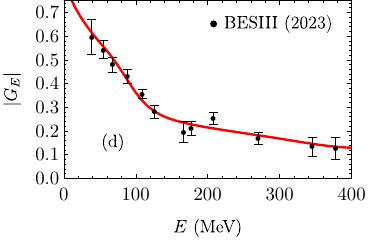}
		\par\end{centering}
	\caption{Energy dependence of the cross section $e^{+}e^{-}\to\Lambda_{c}\bar{\Lambda}_{c}$~ (a),
		the ratio of the electromagnetic form factors $\left|G_{E}/G_{M}\right|$~ (b),
		the absolute value of the magnetic form factor $\left|G_{M}\right|$~ (c)
		and the electric form factor $\left|G_{E}\right|$~ (d). Experimental
		data are taken from Refs.~\cite{Ablikim2018b,Ablikim2023Measurement}.}\label{fig:LambdaC}
\end{figure}

Fig.\ref{fig:LambdaC} shows a comparison of our
calculations with the BESIII experimental data. Plot~(a) shows
the energy dependence of the cross section for the process $e^{+}e^{-}\to\Lambda_{c}\bar{\Lambda}_{c}$.
Due to the influence of the $D$ wave, the energy dependence of the cross section is more
complex than, for example, for the process $e^{+}e^{-}\to\Lambda\bar{\Lambda}$
(cf.~Fig.\ref{fig:Lambda}). In the energy range below $100\,\mbox{Me}$,
the cross section becomes almost constant, and with increasing energy, a sharper drop in the cross section begins. Fig.\ref{fig:LambdaC}(b) shows the energy dependence of the ratio of the electromagnetic form factors for the $\Lambda_{c}$ baryon in the time-like region. Even at energies around $50\mbox{MeV}$ above the threshold, this ratio differs significantly from unity. As mentioned above, the difference between the ratio $\left|G_{E}/G_{M}\right|$ and unity is direct evidence that tensor forces and the contribution of the $D$ wave must be taken into account to correctly describe the interaction between $\Lambda_{c}$ and $\bar{\Lambda}_{c}$. In Ref.~\cite{Ablikim2023Measurement}, the authors emphasize the oscillatory nature of the energy dependence of
$\left|G_{E}/G_{M}\right|$. However, our results show that a fairly good description of these experimental data can be obtained even without pronounced oscillations. Plots (c) and (d) show the energy dependence of the absolute values of the magnetic and electric form factors of the $\Lambda_{c}$ baryon, respectively.

Our model predicts the existence of a bound state of $\Lambda_{c}$ and $\bar{\Lambda}_{c}$ at an energy of approximately $40\,\mbox{MeV}$ below the threshold, which corresponds to $\sqrt{s}\approx4530\,{\text{MeV}}$.
Such a state, if it indeed exists, could manifest itself in other processes, leading to sharp jumps in the cross sections near this energy.

\begin{figure}
	\centering
	\begin{centering}
		\includegraphics[totalheight=5.5cm]{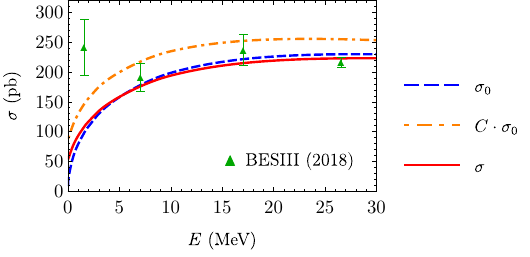}
		\par\end{centering}
	\caption{Effect of the Coulomb interaction on the energy dependence of the cross section
		of the $e^{+}e^{-}\to\Lambda_{c}\bar{\Lambda}_{c}$ process near the threshold.
		The solid line shows the exact cross section $\sigma$ calculated using
		Eq.(\ref{eq:lamc:sig}) with the potential~(\ref{eq:lamc:pot}).
		The dashed line shows the cross section $\sigma_{0}$ calculated using  Eq.(\ref{eq:lamc:sig}),
		but without the Coulomb potential. The dash-dotted line shows
		the cross section $\sigma_{0}$ multiplied by the factor~$C$~(\ref{eq:lamc:Somm}).
		Experimental data are taken from Ref.~\cite{Ablikim2018b}.}\label{fig:Sommer}
\end{figure}

A new effect, compared to the $e^{+}e^{-}\to\Lambda\bar{\Lambda}$ process,
discussed in Sec.\ref{sec:Lambda}, is that the cross section
of the $e^{+}e^{-}\to\Lambda_{c}\bar{\Lambda}_{c}$ process does not vanish
at the threshold (see~Fig.\ref{fig:LambdaC}(a)). This is due to the influence
of the Coulomb interaction between charged baryons. It is important to note, however, that the influence of the Coulomb interaction is not reduced to multiplying the cross section calculated without taking into account the Coulomb attraction, $\sigma_{0}$,
by the Gamow-Sommerfeld-Sakharov factor~\cite{Gamow1928Zur,Gamow1929Zur,Zommerfeld1956Stroenie2,Saharov1948Vzaimodeystvie}, which for the $S$-wave is
\begin{equation}
C=\frac{2\pi\eta}{1-e^{-2\pi\eta}}\,.\label{eq:lamc:Somm}
\end{equation}
Indeed, the Coulomb potential is one of the contributions to the interaction potential (\ref{eq:lamc:pot}) between $\Lambda_{c}$ and $\bar{\Lambda}_{c}$,
along with the strong interaction potential. It is their combined inclusion that yields correct predictions of the cross section behavior near the threshold.
A comparison of the exact cross section calculated using Eq.(\ref{eq:lamc:sig})
with the potential (\ref{eq:lamc:pot}), with the cross section $\sigma_{0}$ calculated
without taking the Coulomb potential into account, and the cross section $C\sigma_{0}$ calculated using the Gamow-Sommerfeld-Sakharov factor (\ref{eq:lamc:Somm}),
is shown in Fig.\ref{fig:Sommer}. It is clear that in some cases
using the factor~$C$ only increases the discrepancy with the exact cross
section.

\section{Processes $e^{+}e^{-}\to p\bar{p}$ and $e^{+}e^{-}\to n\bar{n}$}

This section is devoted to the description of nucleon-antinucleon interactions
within the framework of our approach. Certainly, nucleon-nucleon and nucleon-antinucleon interactions are the most studied of all hadronic interactions.
There is a vast amount of experimental data on both nucleon-nucleon and antinucleon scattering and the production of nucleon-antinucleon
pairs in various processes. Many processes exhibit effects associated with nucleon-antinucleon interactions in both the final and intermediate states. For example, a significant enhancement of the cross sections near the threshold
was found in the processes $e^{+}e^{-}\to p\bar{p}$~\cite{Castellano1973,Delcourt1979,Bisello1983,Bisello1990,Aubert2006,Lees2013,Ablikim2015,Akhmetshin2016,Akhmetshin2019,Ablikim2019,Ablikim2020,Ablikim2021b}
and $e^{+}e^{-}\to n\bar{n}$~\cite{Antonelli1998,Achasov2014,Ablikim2021f,Achasov2022,Ablikim2023Measurements}.
Moreover, in both of these processes, a nontrivial energy dependence is observed on the ratios of the electromagnetic form factors of nucleons. Also, in some processes involving the production of light mesons, a sharp energy dependence of the cross section is observed near the nucleon-antinucleon pair production threshold. ~\cite{Aubert2006a, Akhmetshin2013, Lukin2015, Akhmetshin2019, Aubert2005, Aubert2007c}.
We will show that the interaction of a nucleon with an antinucleon in the final or intermediate state explains the observed effects well.

As mentioned above in the example of the $e^{+}e^{-}\to\Lambda\bar{\Lambda}$ process,
the quantum numbers of the baryon-antibaryon pair produced in the $e^{+}e^{-}$ annihilation process from a virtual photon can have the following values:
total spin $S=1$, total angular momentum $J=1$, orbital angular momentum $L=0$
or $L=2$. However, an important difference between nucleons and the $\Lambda$ hyperon
or $\Lambda_{c}$ baryon is the value of their isotopic spin.
As is known, the proton and neutron have isospin $I=1/2$, with $I_{z}=+1/2$ for the proton and $I_{z}=-1/2$ for the neutron. Therefore, a nucleon-antinucleon
pair produced in $e^{+}e^{-}$ annihilation can exist in
two isospin states, namely
\begin{equation}
\left|I=0,I_{z}=0\right\rangle =\frac{\left|p\bar{p}\right\rangle +\left|n\bar{n}\right\rangle }{\sqrt{2}}\,,\qquad\left|I=1,I_{z}=0\right\rangle =\frac{\left|p\bar{p}\right\rangle -\left|n\bar{n}\right\rangle }{\sqrt{2}}\,.
\end{equation}

Since isotopic spin is not conserved in electromagnetic interactions,
the annihilation of an electron-positron pair produces a nucleon-antinucleon pair  in a superposition of states with $I=0$ and $I=1$. Also,
due to the difference in the masses of the proton and neutron, the thresholds for the production of the $p\bar{p}$ and $n\bar{n}$ states differ by $2.6\,{\text{MeV}}$, which is important to consider when describing the latest precision experimental data obtained near these thresholds. Finally, to account for the large cross section
for the annihilation of the $p\bar{p}$ and $n\bar{n}$ states into light mesons, it is necessary to use the optical potential of the nucleon-antinucleon interaction,
containing an imaginary part.

Early theoretical studies of final-state nucleon-antinucleon interactions used simple models expressing the process cross sections in terms of scattering lengths
\cite{Haidenbauer2006a,Dalkarov2010} or using the Breit-Wigner parametrization. Some studies also calculated the $p\bar {p}$ and $n\bar {n}$ production cross sections using different nucleon-antinucleon interaction potentials~\cite{Haidenbauer2014,VandeWiele2015,Yang2024Study,Jia2025Coupledchannel,Ji2026Understanding}.   Our approach to describing the processes $e^{+}e^{-}\to p\bar{p}$ and $e^{+}e^{-}\to n\bar{n}$ was developed in  Refs.~\cite{Dmitriev2014Isoscalara,Dmitriev2016,Milstein2018,Milstein2022c}
and earlier. We will not present all the formulas here, since
they are given in Ref.~\cite{Milstein2022c}, and they can be obtained similarly
to the formulas in sections~\ref{sec:Lambda} and~\ref{sec:LambdaC}. We note only that to fully account for the Coulomb interaction, tensor forces, and the difference in mass between the proton and neutron, it is necessary to solve the system of Schr\"odinger equations for four coupled channels corresponding to the states
$p\bar{p}$ and $n\bar{n}$ with orbital angular momenta $L=0$ and $L=2$.

As discussed in previous sections, different parametrization of the interaction potential can be chosen to describe the near-threshold behavior of process cross sections and hadron form factors. With a suitable choice of potential parameter values, different models will yield similar predictions for experimentally observed quantities. Recall that this is due to the fact that the characteristic size of the wave function of a pair of hadrons with low relative velocities is large compared to the characteristic interaction radius. In this case, the details of the interaction between the hadrons do not play a decisive role. It turns out that currently available full-fledged models of nucleon-antinucleon interactions (see, for example, Refs.~\cite{el-bennich2009paris,zhou2012energy,Dai2017}), used to describe $N\bar{N}$ scattering, are, for various reasons, poorly suited to calculating wave functions at $r\to0$. Therefore, in Refs.~\cite{Dmitriev2016,Milstein2018,Milstein2022c}, we proposed our own simple model of nucleon-antinucleon interaction, which is limited to describing states with spin $S=1$ and total angular momentum
$J=1$, since only $N\bar{N}$ pairs with such quantum numbers are produced in the process of $e^{+}e^{-}$ annihilation. Our model of the strong interaction
includes the long-range contribution of pion exchange, as well as the potentials
of isoscalar and isovector exchange at small distances, describing the interaction in the $S$-wave, $D$-wave, and tensor forces. Small-range
potentials contain real and imaginary parts and are parameterized by
rectangular potential wells (for details, see Ref.~\cite{Milstein2022c}).

We select the parameters of our model to best describe the available experimental data obtained in the study of nucleon-antinucleon scattering and the production of nucleon-antinucleon pairs in $e^{+}e^{-}$ annihilation. First, we use the results of the analysis of nucleon-antinucleon scattering data performed by the Nijmegen group~\cite{zhou2012energy},
which determined the partial cross sections for proton-antiproton scattering,
as well as the charge exchange process $p\bar{p}\to n\bar{n}$. Second,
we consider data for the cross sections of the $e^{+}e^{-}\to p\bar{p}$
and $e^{+}e^{-}\to n\bar{n}$ processes, obtained in different years at the BaBar~\cite{Lees2013}, CMD-3~\cite{Akhmetshin2016,Akhmetshin2019},
SND~\cite{Achasov2022}, and BESIII~\cite{Ablikim2020,Ablikim2021b,Ablikim2021f} detectors.
Third, we describe the available experimental data for the electromagnetic
form factors of the proton and neutron. The optimal parameters of our model,
obtained by minimizing $\chi^{2}$, are given in Ref.\cite{Milstein2022c}.
For these parameter values, we obtained the value $\chi^{2}/N_{\mathrm{df}}=105{.}6/89=1{.}19$.

\begin{figure}[!tb]
	\centering
	\begin{centering}
		\includegraphics[width=1\textwidth]{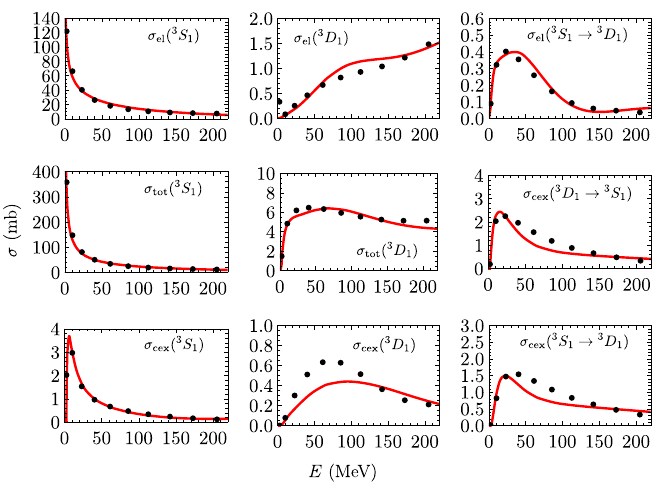}
		\par\end{centering}
	\centering{}\caption{Comparison of our predictions for partial nucleon-antinucleon scattering cross sections with the results of the analysis performed by the Nijmegen group~\cite{zhou2012energy}.}\label{fig:Nijmegen}
\end{figure}

Fig.\ref{fig:Nijmegen} shows a comparison of the partial cross sections
for nucleon-antinucleon scattering obtained within the framework of our model
with the results of the analysis carried out by the Nijmegen group~\cite{zhou2012energy}.
Note that the initial experimental data in Ref.~\cite{zhou2012energy}
were differential and total scattering cross sections, as well as
some other observable quantities. Using these data, the
Nijmegen model of nucleon-antinucleon interaction was constructed, and then
the partial scattering cross sections were calculated. However, there are other
interaction models that also describe the experimental
data well, but predict somewhat different partial cross sections (see, for example, the comparison of different models in Ref.~\cite{Carbonell2023}).
Therefore, the accuracy with which partial cross sections can be extracted from experimental data is difficult to estimate, and the errors in the partial cross sections are not shown in the Nijmegen graphs.
Overall, the model we use reproduces the Nijmegen partial cross sections well, since differences are noticeable only in those partial
waves where the cross sections themselves are small.

\begin{figure}
	\centering
	\includegraphics[totalheight=5.1cm]{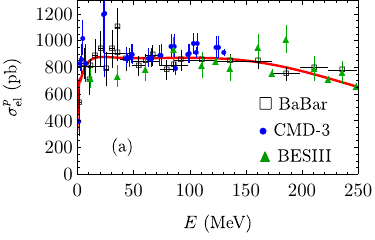}\hfill{}\includegraphics[totalheight=5.1cm]{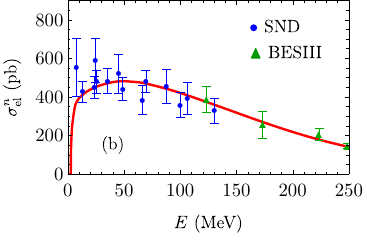}
	
	\includegraphics[totalheight=5.1cm]{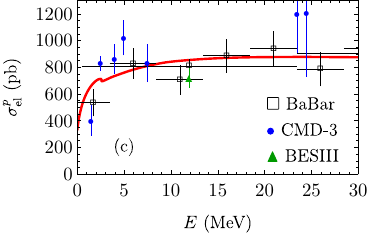}\hfill{}\includegraphics[totalheight=5.1cm]{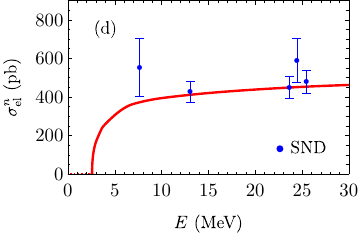}
	
	\caption{Energy dependence of the $p\bar{p}$ (a, c) and $n\bar{n}$ (b, d) production cross sections in $e^{+}e^{-}$ annihilation. The bottom row shows the near-threshold energy region in more detail. Experimental data taken from the BaBar collaborations~\cite{Lees2013}, CMD-3~\cite{Akhmetshin2016,Akhmetshin2019},
		SND~\cite{Achasov2022}, and BESIII~\cite{Ablikim2020,Ablikim2021b,Ablikim2021f}.}\label{fig:NNProd}
\end{figure}

Fig.~\ref{fig:NNProd} shows a comparison of our results
for the cross sections of the $e^{+}e^{-}\to p\bar{p}$ and $e^{+}e^{-}\to n\bar{n}$ processes with experimental data. Here, we do not include some results
from older experiments that clearly contradict new, more accurate
data or have been superseded by them. Overall, our model describes well
the cross sections of $p\bar{p}$ and $n\bar{n}$ production in the energy range under consideration. A characteristic feature of the energy dependence of these cross sections
is an extremely rapid increase near the threshold, reaching the  plateau,
followed by a smooth decline. Such a dependence of the cross sections on
energy is apparently impossible to obtain without taking into account the $D$-wave
contributions. It is precisely by mixing states with orbital momenta
$L=0$ and $L=2$ due to the action of tensor forces that it becomes possible
to reproduce such a nontrivial dependence of cross sections on energy.

\begin{figure}[!tb]
	\centering
	\includegraphics[totalheight=5.1cm]{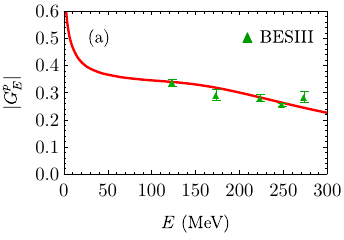}\hfill{}\includegraphics[totalheight=5.1cm]{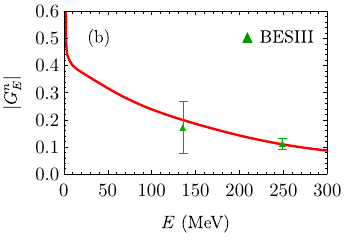}
	
	\includegraphics[totalheight=5.1cm]{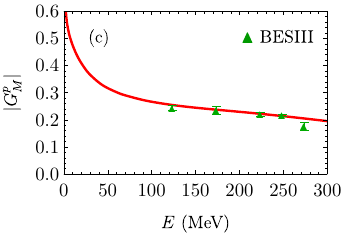}\hfill{}\includegraphics[totalheight=5.1cm]{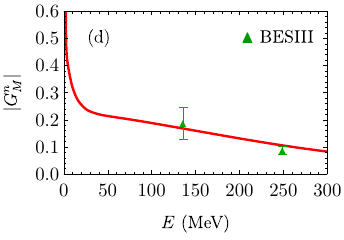}
	
	\includegraphics[totalheight=5.1cm]{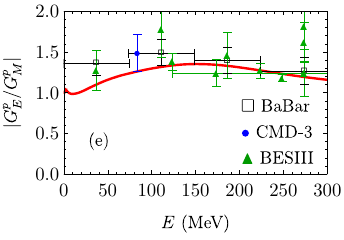}\hfill{}\includegraphics[totalheight=5.1cm]{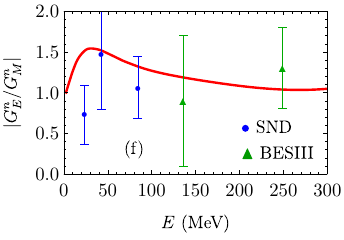}
	
	\caption{Energy dependence of the electromagnetic form factors of proton (a, c)
		and neutron (b, d), as well as the ratios $\left|G_{E}/G_{M}\right|$
		(d, e). Experimental data are taken from the Refs.~
		\cite{Lees2013, Akhmetshin2016, Ablikim2019, Ablikim2020, Ablikim2021b, Achasov2022, Ablikim2023Measurements}}.\label{fig:NNGeGm}
\end{figure}

Fig.~\ref{fig:NNGeGm} shows a comparison with experimental
data of our results for electromagnetic form factors, as well as the ratios $\left|G_{E}/G_{M}\right|$ for the proton and neutron.
Here, good agreement with the experimental results was also obtained.
We again emphasize that the difference from unity in the ratio
of electromagnetic form factors is a direct consequence of the mixing of the $S$-wave and $D$-wave due to the action of tensor forces.

The approach considered can also be applied to take into account the nucleon-antinucleon interaction in the final state in other processes. For example,
an enhancement of decay probabilities for the invariant mass of the $p\bar{p}$ pair
near two proton masses was observed in the decays of $J/\psi\to p\bar{p}\gamma\left(\omega,\pi^{0},\eta\right)$
and $\psi(2S)\to p\bar{p}\gamma\left(\pi^{0},\eta\right)$~\cite{Bai2001,Bai2003,Ablikim2008,Ablikim2009,Alexander2010,Ablikim2012,Ablikim2013b,Ablikim2013,Ablikim2013a,Ablikim2024Measurement,Ablikim2025Partial}.
In Refs.~\cite{Dmitriev2016a, Milstein2017, Salnikov2024Meson}, it was shown that the proton-antiproton interaction in the final state explains the distributions of the invariant mass $p\bar{p}$ observed in these processes.

\section{$N\bar{N}$ interaction in the intermediate state}

Nucleon-antinucleon interaction manifests not only in processes involving direct nucleon-antinucleon pair production, but also in some other processes. For example, in some processes involving the production of multi-mezonic states in electron-positron annihilation, a sharp drop in cross sections was observed at the threshold of nucleon-antinucleon pair production. A natural explanation for such effects is the interaction of virtual nucleon-antinucleon pairs in an intermediate state. In Refs.~\cite{Haidenbauer2015, Ji2026Understanding}
using chiral models~\cite{Kang2014, Dai2017}, it was also shown
that the $N\bar{N}$ interaction can explain the behavior
of the cross sections of the processes $e^{+}e^{-}\to6\pi$, $e^{+}e^{-}\to5\pi$, $e^{+}e^{-}\to\omega\pi^{+}\pi^{-}\pi^{0}$
and $e^{+}e^{-}\to\pi^{+}\pi^{-}K^{+}K^{-}$ near the threshold of production
of nucleon-antinucleon pairs.

In Sec.~\ref{sec:inelastic}, we already discussed that our approach can describe processes in which a virtual hadron pair is produced, which then annihilates into other final states.
The interaction of virtual hadrons in an intermediate state can lead to a nontrivial energy dependence of the cross sections of such processes.
Here, we will examine the effects of interaction in the intermediate state in more detail, using the example of the interaction of virtual nucleon-antinucleon
pairs. Recall that we call the cross section for the production of real nucleon-antinucleon
pairs in the final state the elastic cross section, and the cross section of processes involving the production of virtual $N\bar{N}$ pairs in the intermediate
state and their subsequent annihilation into light mesons the inelastic cross section. The total cross section for the production of $N\bar{N}$ pairs is the sum of the elastic and inelastic cross sections.

\begin{figure}
	\centering
	\includegraphics[totalheight=5.1cm]{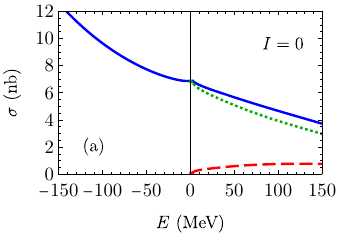}\hfill{}\includegraphics[totalheight=5.1cm]{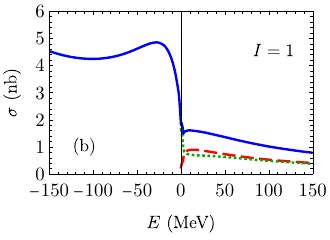}
	
	\caption{Energy dependences of elastic (dashed lines), inelastic (dotted
		lines) and total (solid lines) cross sections for the production of $N\bar{N}$ in states
		with isospins $I=0$ (a) and $I=1$ (b) in $e^{+}e^{-}$ annihilation.}\label{fig:NNIso1}
\end{figure}

Effects associated with nucleon-antinucleon interactions in the intermediate
state have been observed in some processes involving the production of multi-mison states in $e^{+}e^{-}$ annihilation. Since such states often
have a specific isospin value (rather than being a mixture of states
with $I=0$ and $I=1$), to predict the cross sections of such processes, we
need contributions to the cross sections associated with the nucleon-antinucleon interaction
in a specific isospin state. According to the results
in Sec.~\ref{sec:inelastic}, the total cross section for hadron pair production
is expressed in terms of their Green's function. A detailed description of the relationship between the total cross sections for $N\bar{N}$ production in isoscalar and isovector states
and the Green's function of the Sch\"odinger equations, as well as a method for calculating it, are given in our papers~\cite{Milstein2018,Milstein2022c}. Our predictions
for the total, elastic, and inelastic contributions to the cross sections for the production of nucleon-antinucleon
pairs with a given isospin are shown in Fig.~\ref{fig:NNIso1}.
Our model of nucleon-antinucleon interactions predicts a noticeable
drop in the total and inelastic cross sections below the threshold for the production of a real $N\bar{N}$ pair in the channel with $I=1$. We note that the drop in the total cross section significantly exceeds the elastic cross section for the production of a real $N\bar{N}$ pair. Since a virtual nucleon-antinucleon
pair in an isovector state can annihilate into
different meson states, this jump of cross section  can manifest itself to varying degrees
in the cross sections of different processes.

\begin{figure}[!tb]
	\centering
	\includegraphics[totalheight=5.1cm]{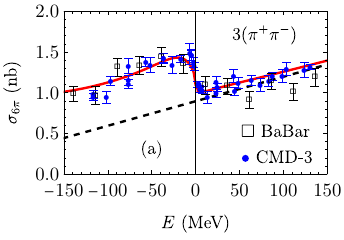}\hfill{}\includegraphics[totalheight=5.1cm]{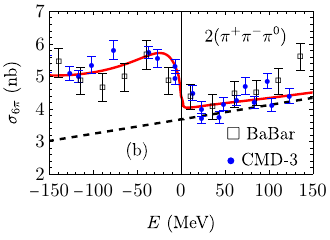}
	\begin{centering}
		\includegraphics[totalheight=5.1cm]{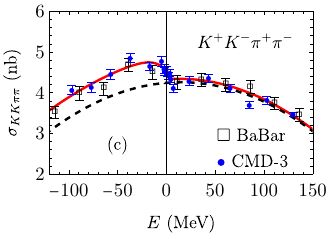}
		\par\end{centering}
	\caption{Energy dependence of process cross sections $e^{+}e^{-}\to3\left(\pi^{+}\pi^{-}\right)$,
		$e^{+}e^{-}\to2\left(\pi^{+}\pi^{-}\pi^{0}\right)$ and $e^{+}e^{-}\to K^{+}K^{-}\pi^{+}\pi^{-}$.
	The dashed lines show contributions to the cross sections of processes not associated with the production of a virtual $N\bar{N}$ pair. Experimental data
	are taken from Refs.~\cite{Aubert2006a,Akhmetshin2013,Akhmetshin2019},
		\cite{Aubert2006a,Lukin2015} and
			Refs.~\cite{Aubert2005,Aubert2007c,Akhmetshin2019}, respectively.}\label{fig:mesons}
\end{figure}

Let us now consider several specific processes in which effects associated with nucleon-antinucleon interactions in the intermediate
state have been observed. The processes $e^{+}e^{-}\to3\left(\pi^{+}\pi^{-}\right)$ and $e^{+}e^{-}\to2\left(\pi^{+}\pi^{-}\pi^{0}\right)$ have been studied in most detail. Since the $G$-parity
of the final states in these processes is $+1$, and the $C$-parity
is $-1$, they can only be formed through the annihilation
of a virtual nucleon-antinucleon pair with $I=1$. Other intermediate
states that do not contain an $N\bar{N}$ pair also contribute to the cross section
of the $e^{+}e^{-}\to6\pi$ process. However, such contributions to the cross section should have a smooth energy dependence in the vicinity of the production threshold
of real $N\bar{N}$ pairs. The probability of annihilation of a virtual pair
$N\bar{N}$ in $6\pi$ is also a smooth function of energy and can be considered constant near the threshold. Therefore, we will describe the energy dependence
of the cross sections of processes $e^{+}e^{-}\to6\pi$ by the expression
\begin{equation}
\sigma_{6\pi}(E)=A\cdot\sigma^{1}_{\text{inel}}(E)+B\cdot E+C\,,\label{eq:nuc:6pi}
\end{equation}
where $A$, $B$, and $C$~ are some energy-independent coefficients. Comparison of Eq.~(\ref{eq:nuc:6pi}) with experimental
data~\cite{Aubert2006a, Akhmetshin2013, Lukin2015, Akhmetshin2019} for the cross sections of the processes $e^{+}e^{-}\to6\pi$ yields the values of the parameters
$A=0.12$, $B=3.2\cdot10^{-3}{\text{nb}/\text{MeV}}$,
$C=0.9\,{\text{nb}}$ for the process $e^{+}e^{-}\to3\left(\pi^{+}\pi^{-}\right)$
and $A=0.5$, $B=4{.}8\cdot10^{-3}{\text{nb}/\text{MeV}}$,
$C=3{.}6\,{\text{nb}}$ for the process $e^{+}e^{-}\to2\left(\pi^{+}\pi^{-}\pi^{0}\right)$.
A comparison of our predictions with experimental results is shown
in Fig.~\ref{fig:mesons}. Overall, our model reproduces well
the sharp drop in cross sections near the production threshold of real nucleon-antinucleon
pairs. The coefficient~$A$ corresponds semantically to the probability that
a virtual nucleon-antinucleon pair with $I=1$ annihilates into a certain
final state. Thus, we see that the total probability
of the formation of a $6\pi$ state from such an $N\bar{N}$ pair is
62\%. Indeed, this value is quite close to the relative
probability of the annihilation of a resting nucleon-antinucleon pair with $I=1$
into six pions, which is about 56\%, see~\cite{Klempt2005}.

Another process in which a noticeable jump in the cross section was also observed at the threshold of production of a real nucleon-antinucleon pair is the process
$e^{+}e^{-}\to K^{+}K^{-}\pi^{+}\pi^{-}$. The final state here
can have either $I=1$ or $I=0$. However, our model predicts
that the contribution to the cross section associated with the nucleon-antinucleon interaction
in the state with $I=0$ does not exhibit sharp jumps near the threshold of production
of a real $N\bar{N}$ pair (see Fig.~~\ref{fig:NNIso1}). Taking into account
also the smooth energy dependence of the contributions to the cross section
associated with other intermediate states, we write the cross section of the process
as
\begin{equation}
\sigma_{KK\pi\pi}(E)=A\cdot\sigma^{1}_{\text{inel}}(E)+B\cdot E^{2}+C\cdot E+D\,,
\end{equation}
where $A$, $B$, $C$, and $D$ are some energy-independent
coefficients. Here, only the term containing $\sigma^{1}_{\text{inel}}$ has a jump at the threshold for the production of a real $N\bar{N}$ pair, while the remaining
terms describe a set of contributions to the cross section that are weakly dependent
on energy. By fitting the experimental data~\cite{Aubert2005,Aubert2007c,Akhmetshin2019}
we find $A=0.12$, $B=-6.6\cdot10^{-5}\,{\text{nb}/\text{MeV}^{2}}$,
$C=2\cdot10^{-3}\text{nb}/\text{MeV}$, $D=4.2\,{\text{nb}}$.
A comparison of our predictions for the cross section of the process $e^{+}e^{-}\to K^{+}K^{-}\pi^{+}\pi^{-}$
with the experimental results is shown in Fig.~\ref{fig:mesons}(c).

Thus, within the framework of our nucleon-antinucleon interaction model, we can describe not only the energy dependence of the cross sections for the production of real $p\bar{p}$ and $n\bar{n}$ pairs, but also obtain the contributions to the cross sections of other processes associated with the interaction of virtual $N\bar{N}$ pairs. We expect that a sharp drop in the cross sections at the threshold of production of real $N\bar{N}$ pairs can also be observed in other processes involving the production of mesons in isovector states. However, we cannot predict in advance which processes should be searched for manifestations of nucleon-antinucleon interaction in the intermediate state. At the same time, the cross sections for the production of
isoscalar states should not exhibit significant features in this energy range (see Fig.~ \ref{fig:NNIso1}).

\section{Production of $D^{(*)}\bar{D}^{(*)}$ in $e^{+}e^{-}$ annihilation}\label{sec:DD}

In this section, we consider the final-state interaction in
processes with $D^{(*)}\bar{D}^{(*)}$ pair production in $e^{+}e^{-}$ annihilation. The complex energy dependence of the cross sections near the thresholds
in processes with $D^{(*)}$ meson pair production in $e^{+}e^{-}$ annihilation
has been observed at the BaBar~\cite{Aubert2007Study}, Belle~\cite{Pakhlova2007Measurement,Pakhlova2008Measurement,Zhukova2018Angular},
CLEO~\cite{Cronin-Hennessy2009Measurement}, and BESIII~\cite{Ablikim2022Cross,Ablikim2024Precise} detectors.
The mass difference between vector and pseudoscalar $D$ mesons is approximately $140\,\mbox{ MeV}$, which is significantly smaller than the energy range studied. Therefore, accounting for the mixing of the $D\bar{D}$, $D\bar{D}^{*}$ and $D^{*}\bar{D}^{*}$ states is important for describing the energy dependence of the production cross sections for these states. Several groups have proposed different approaches to describing the interaction between $D^{(*)}$ mesons~\cite {Husken2024Polesa,Ye2025Resonance}.
Below, we consider the final-state interaction within the framework of our approach. In principle, at the energies at which $D^{(*)}\bar{D}^{(*)}$ pairs are produced, a pair of $D_{s}\bar{D}_{s}$ mesons can also be produced.
However, we expect that, due to the presence of $s$ quarks, the $D_{s}\bar{D}_{s}$ state will weakly mix with $D^{(*)}\bar{D}^{(*)}$ states. Therefore, in our analysis, we restrict ourselves to processes involving the production of
$D^{(*)}\bar{D}^{(*)}$ pairs, for which mixing between different
reaction channels plays a significant role.

Since the intrinsic parity of bot h$D$ and $D^{*}$ mesons is negative, the $P$ parity of the meson-anti-meson state is related to their relative orbital angular momentum $L$ by $P=\left(-1\right)^{L}$.
Thus, in the $e^{+}e^{-}\to D^{(*)}\bar{D}^{(*)}$ processes, only pairs with odd $L$ can be produced from a virtual photon,
and near the thresholds, the contribution of states with $L = 1$ should dominate. In the case of $D\bar{D}$ pair production, the spin of the final state is $S = 0$, and the $C$ parity is $C = \left(-1\right)^{L+S} = -1$. The $C$-odd state of a pseudoscalar
and vector meson has spin $S=1$ and corresponds to the wave function
$(D\bar{D}^{*}+D^{*}\bar{D})/\sqrt{2}$. The $D^{*}\bar{D}^{*}$ pair in a $C$-odd state, in principle, can have either spin $S=0$ or $S=2$. However, experimental data that would allow us
to distinguish between these spin states are currently lacking,
therefore, in what follows, we will assume some averaged
state of the $D^{*}\bar{D}^{*}$ pair.

Production of a pair of $D^{(*)}$ mesons in $e^{+}e^{-}$ annihilation occurs as follows. At small distances $r\sim1/\sqrt{s}$, a quark-antiquark pair $c\bar{c}$ with isospin
$I=0$ is produced from a virtual
photon. Then, as the $c$ quarks fly apart, hadronization occurs
and $D^{(*)}$ mesons are formed. However, due to the difference in the masses of charged
and neutral $D^{(*)}$ mesons, as well as the Coulomb forces between charged
mesons, the final-state interaction violates isotopic
invariance. Thus, to fully describe the final-state interaction in processes with $D^{(*)}\bar{D}^{(*)}$ pair production, we must consider six different final states (three with charged mesons and three with neutral mesons).

The six-channel radial wave function of the $D^{(*)}$-mesons system satisfies a system of radial Schr\"odinger equations with orbital
momentum $L=1$ (details are given in Ref.~\cite{Salnikov2024Production}).
As has been repeatedly discussed above, accounting for final-state interactions in processes with hadron pair production near the threshold can be reduced to a small number of parameters. The details of the interactions between hadrons in such cases do not play a decisive role. Therefore, we can choose the most convenient parametrization of the interaction potentials for calculations.
As a simple parametrization of the potential matrix, we choose isoscalar and isovector exchange potentials in the form of rectangular wells corresponding to the interaction in each channel, as well as the transitions between these channels. The cross sections for the production of $D^{(*)}\bar{D}^{(*)}$ pairs
are expressed in terms of the derivatives of the radial wave functions at zero, since
these pairs are produced in states with $L=1$, and also in terms of the constants
$g_{i}$ associated with the probabilities of the production of various $D^{(*)}$ meson pairs
at small distances (see Ref.~\cite{Salnikov2024Production}).

The cross sections of the $e^{+}e^{-}\to D^{(*)}\bar{D}^{(*)}$ processes in the near-threshold
energy region were measured at the BaBar~\cite{Aubert2007Study},
Belle~\cite{Pakhlova2008Measurement,Zhukova2018Angular}, CLEO~\cite{Cronin-Hennessy2009Measurement}
and BESIII~\cite{Ablikim2022Cross,Ablikim2024Precise} detectors. In addition, raw measurement results for the cross section of the $e^{+}e^{-}\to D\bar{D}$ process at BESIII in the region of the $\psi(3770)$ resonance
were presented in the dissertation paper~\cite{Julin2017}. Although
these data were not officially published in the collaborative paper,
in Ref.~\cite{Husken2024Polesa}, the Born cross section was calculated from these measurements
(taking into account corrections for
vacuum polarization and photon emission in the initial state). All
of the above measurements were used by us to determine the parameters
of the interaction potentials in the $D^{(*)}$ meson system (see Ref.~\cite{Salnikov2024Production}
and Ref.~\cite{Salnikov2026Priporogovye}).

We calculated the value of $\chi^{2}$, which indicates how well our model describes all the experimental data for the cross sections of the processes $e^{+}e^{-}\to D^{(*)}\bar{D}^{(*)}$. We then found the parameters of the potentials, as well as the constants $g_{i}$, to obtain the smallest value of $\chi^{2}$. The fitting resulted in the value $\chi^{2}/N_{\mathrm{df}}=397/338=1.18$ for 367 experimental
points. The parameters of the potentials found, taking into account experimental
data in the vicinity of the $\psi(3770)$ resonance~\cite{Husken2024Polesa}
are presented in Ref.~\cite{Salnikov2026Priporogovye}.

\begin{figure}[!tb]
	\centering
	\begin{centering}
		\includegraphics[totalheight=5.5cm]{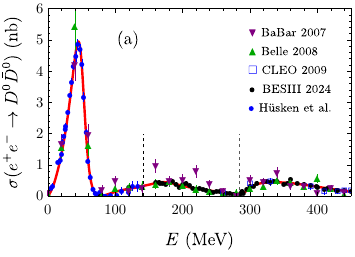}\hfill{}\includegraphics[totalheight=5.5cm]{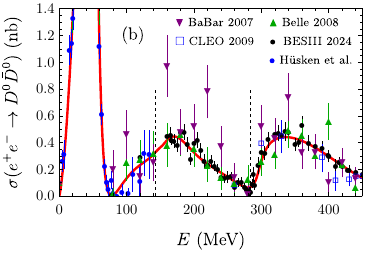}
		\par\end{centering}
	\begin{centering}
		\includegraphics[totalheight=5.5cm]{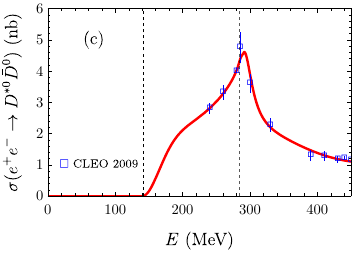}\hfill{}\includegraphics[totalheight=5.5cm]{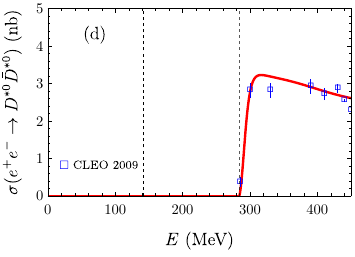}
		\par\end{centering}
	\centering{}\caption{Comparison with experimental data of our predictions for the energy dependence of the cross sections for pair production of neutral $D^{(*)}$ mesons. Experimental
		data are taken from Refs.~\cite{Aubert2007Study,Pakhlova2008Measurement,Cronin-Hennessy2009Measurement,Dong2018Derived,Ablikim2024Precise,Husken2024Polesa}.
		The energy is measured from the production threshold of $D^0\bar{D}^0$, and the vertical dashed lines show the production thresholds $D^{*0}\bar{D}^{0}$
		and $D^{*0}\bar{D}^{*0}$.}\label{fig:D0D0}
\end{figure}

\begin{figure}[!tb]
	\centering
	\begin{centering}
		\includegraphics[totalheight=5.5cm]{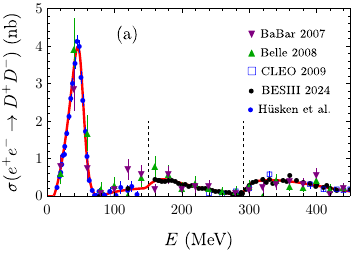}\hfill{}\includegraphics[totalheight=5.5cm]{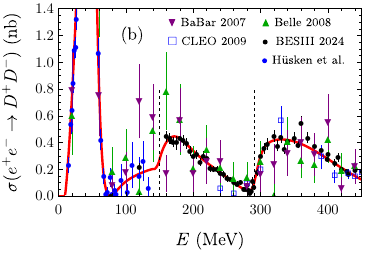}
		\par\end{centering}
	\begin{centering}
		\includegraphics[totalheight=5.5cm]{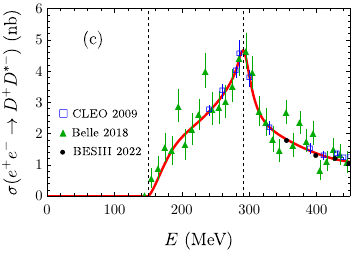}\hfill{}\includegraphics[totalheight=5.5cm]{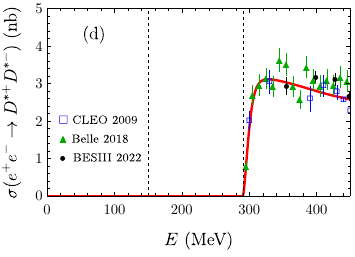}
		\par\end{centering}
	\centering{}\caption{Comparison with experimental data of our predictions for the energy dependence of the cross sections for the production of charged $D^{(*)}$ meson pairs. Experimental
		data are taken from Ref.~\cite{Aubert2007Study,Pakhlova2008Measurement,Cronin-Hennessy2009Measurement,Dong2018Derived,Zhukova2018Angular,Ablikim2022Cross,Ablikim2024Precise,Husken2024Polesa}.
		The energy is measured from the birth threshold $D^0\bar{D}^0$, and the vertical dashed lines show the production thresholds $D^{+}D^{*-}$
		and $D^{*+}D^{*-}$.}\label{fig:DD}
\end{figure}

Comparisons of our calculations with experimental data
are shown in Fig.~\ref{fig:D0D0} for the cross sections for the production of pairs of neutral
$D^{(*)}$ mesons and in Fig. ~\ref{fig:DD} for the cross sections for the production
of pairs of charged $D^{(*)}$ mesons. Our model describes well the nontrivial
energy dependence of the cross sections of all these processes. In particular,
we are able to reproduce deep dips in the cross sections for the production of $D^{0}\bar{D}^{0}$
and $D^{+}D^{-}$ at energies of about $80\,\mbox{MeV}$ and about $280\,\mbox{MeV}$. While
the second dip falls precisely at the $D^{*}\bar{D}^{*}$ production threshold,
the first dip lies slightly below the $D^{*}\bar{D}$ production threshold. This complex energy dependence of the cross sections is a consequence of the interaction
between different reaction channels and the interference of different contributions to the
cross sections. Simultaneously taking into account all contributions to the interaction potential
(both diagonal and off-diagonal) is important for describing the available experimental data. For the cross section ratio $\sigma(e^{+}e^{-}\to D^{+}D^{-})/\sigma(e^{+}e^{-}\to D^{0}\bar{D}^{0})$
near the maximum of the $\psi(3770)$ resonance cross section, our calculations yield a value of $0.8$, which also corresponds to measurements from the CLEO detector~\cite{Bonvicini2014Updated}.

The $\psi(3770)$ resonance in the $D\bar{D}$ production cross section in our approach arises due to the presence of a virtual level in the system of interacting $D$-mesons in our multichannel problem. This result may seem to contradict the generally accepted idea that $\psi(3770)$ represents a bound state of $c\bar{c}$.
However, it should be kept in mind that the exact $\psi(3770)$ wave function
contains not only the contribution of the $c\bar{c}$ state but also contributions from other
states to which a transition from $c\bar{c}$ is possible due to interactions.
When calculating various matrix elements, the dominant contribution will come from different components of the total $\psi(3770)$ wave function. In particular, the main contribution to the matrix element of the transition to the final state $D\bar{D}$ comes from the $\psi(3770)$ component of the wave function, which contains an admixture of light quarks. The interaction in this state is described within our approach by introducing a certain effective potential for the interaction between $D$ mesons.

\section{Production of $B^{(*)}\bar{B}^{(*)}$ in $e^{+}e^{-}$ annihilation}

In this section, we consider the influence of final-state interactions
on the cross sections of $e^{+}e^{-}\to B^{(*)}\bar{B}^{(*)}$ processes near the thresholds. The BaBar detector measured the total cross section for the production
of states containing the $b\bar{b}$ quark pair~\cite{Aubert2009}.
This cross section, in the energy range under consideration, corresponds to the sum of the cross sections for the production of $B\bar{B}$, $B^{*}\bar{B}$, and $B^{*}\bar{B}^{*}$. In addition, exclusive cross sections for the production of $B\bar{B}$, $B^{*}\bar{B}$, and $B^{*}\bar{B}^{*}$ were measured at the Belle and Belle~II detectors~\cite{Mizuk2021,Adachi2024Measurement}.
The energy dependence of these cross sections revealed several complex peaks and dips, which may be manifestations of final-state interactions in the multichannel problem.
Indeed, the threshold for $B^{*}\bar{B}$ production lies $45\,\mbox{MeV}$ above the threshold for $B\bar{B}$, and the threshold for $B^{*}\bar{B}^{*}$ production lies $90\,\mbox{MeV}$ above the threshold for $B\bar{B}$. Since all three final states
are produced in $e^{+}e^{-}$ annihilation via a single virtual photon,
they have identical quantum numbers $J^{PC}=1^{--}$. This means that in the
process of $B^{(*)}\bar{B}^{(*)}$ meson pair production, mixing
of different final states and transitions between them is possible. It is the coupling between different reaction channels in such processes that leads to nontrivial
near-threshold effects.

Since the $b\bar{b}$ state produced at small distances is
isoscalar, and the effects of isotopic invariance violation are quite weak, the admixture of the isovector state arising from the $B^{(*)}$-meson interaction is small. This means that the interaction potential in the isovector channel has a weak effect on the production cross sections of $B^{(*)}\bar{B}^{(*)}$ pairs, summed over charge states.
Therefore, the available experimental data allow us to reliably determine the parameters of the interaction potentials between $B^{(*)}$-mesons in isoscalar
states. To determine the parameters of the interaction potential in
isovector channels, experimental data for the exclusive
production cross sections of charged and neutral $B^{(*)}$-mesons are needed. Unfortunately, the experiments~\cite{Aubert2009, Mizuk2021, Adachi2024Measurement}
did not distinguish between the production of pairs of charged and neutral $B^{(*)}$ mesons.
In other studies, the ratios of the probabilities of the decay of $\Upsilon(4S)$ into $B^{+}B^{-}$ and $B^{0}\bar{B}^{0}$ pairs at the resonance peak were measured~\cite{Alexander2001,Aubert2002,Athar2002,Hastings2003,Aubert2004a,Choudhury2023}, and the ratio of the cross sections for the production of $B^{+}B^{-}$ and $B^{0}\bar{B}^{0}$ pairs in the vicinity of $\Upsilon(4S)$ were measured~\cite{Abumusabh2026Measurements}.
However, these data are insufficient to fully determine the contributions
of isoscalar and isovector exchanges to the $B^{(*)}\bar{B}^{(*)}$ interaction potential. Therefore, we used a somewhat simpler interaction model than for $D^{(*)}$ mesons (details are given in Ref.~\cite{Salnikov2024, Salnikov2026Charge}). In Ref.~\cite{Salnikov2026Charge}
we proposed three sets of parameters that describe well the available experimental data for the total cross sections. However, these three models give different predictions for the production cross sections of charged and neutral $B^{(*)}$ meson pairs
above $\Upsilon(4S)$, where experimental data are not yet available.

\begin{figure}[!tb]
	\centering
	\begin{centering}
		\includegraphics[totalheight=5.5cm]{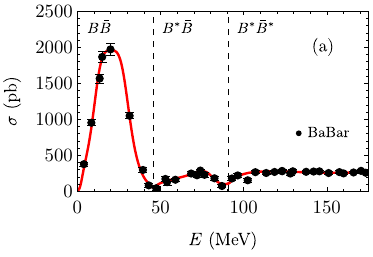}\hspace*{\fill}\includegraphics[totalheight=5.5cm]{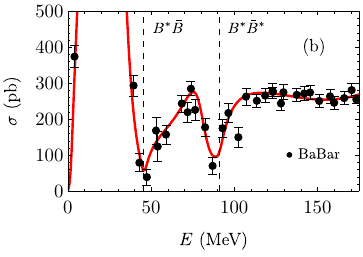}
		\par\end{centering}
	\begin{centering}
		\includegraphics[totalheight=5.5cm]{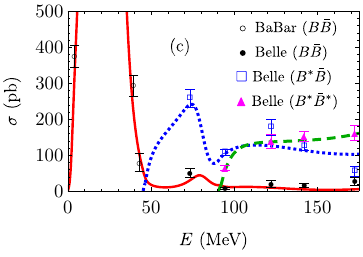}
		\par\end{centering}
	\caption{Energy dependence of our predictions for the sum of the cross sections for the production of
		$B\bar{B}$, $B^{*}\bar{B}$, and $B^{*}\bar{B}^{*}$ in $e^{+}e^{-}$
		annihilations (a, b). Energy dependence of our predictions for the
		exclusive cross sections (c) for the processes $e^{+}e^{-}\to B\bar{B}$ (solid
		line), $e^{+}e^{-}\to B\bar{B}^{*}$ (dotted line), and $e^{+}e^{-}\to B^{*}\bar{B}^{*}$
		(dashed line). Experimental data for the total cross section
		are taken from Dong2020, and for the exclusive cross sections from Refs.~\cite{Mizuk2021, Adachi2024Measurement}.
		The vertical dashed lines indicate the thresholds for the production of the states
		$B^{*}\bar{B}$ and $B^{*}\bar{B}^{*}$.}\label{fig:BB}
\end{figure}

To determine the optimal parameters of the interaction potential in the $B^{(*)}$-meson system, we minimize the value of $\chi^{2}$ (the sum
of the squares of the standard deviations from the experimental data), calculated
using all available experimental data. The potential parameters obtained as a result of the fitting for all three model variants are presented in the Ref.~\cite{Salnikov2026Charge}. The final value is $\chi^{2}/N_{\mathrm{df}}=50.2/32=1.57$
for 49 experimental points corresponding to the cross sections of the processes
$e^{+}e^{-}\to B^{(*)}\bar{B}^{(*)}$. A comparison of the calculated results
with experimental data for the $B\bar{B}$, $B^{*}\bar{B}$
and $B^{*}\bar{B}^{*}$ production cross sections is shown in Fig.~\ref{fig:BB}. Note that for all three model variants, the $B^{(*)}$-meson pair production cross sections,
summed over charge states, are almost identical due to the weak sensitivity of these cross sections to the effects of isotopic invariance violation.

A characteristic feature of the behavior of the total $e^{+}e^{-}\to b\bar{b}$ cross section~\cite{Dong2020}
is the presence of deep dips at the production thresholds of
$B^{*}\bar{B}$ and $B^{*}\bar{B}^{*}$. In our model
such dips arise due to the possibility of transitions between different
states of the $B^{(*)}$ meson pair ($B\bar{B}$, $B^{*}\bar{B}$
and $B^{*}\bar{B}^{*}$). This results in interference of several amplitudes in the cross sections, which leads to a nontrivial energy dependence of the cross sections. The coupling between different reaction channels can also lead to other interesting effects. For example, if we set all potentials mixing different channels to zero, then in our model there exists a bound state in the $B^{*}\bar{B}^{*}$ system at an energy of $65\,\mbox{MeV}$ (that is, $25\,\mbox{MeV}$ below the threshold). However, taking into account possible transitions between channels, this state is no longer bound, since it can decay into $B\bar{B}$ or $B^{*}\bar{B}$. This leads to a peak appearing in the $B\bar{B}$ and $B^{*}\bar{B}$ channels at an energy of about $75\,\mbox{MeV}$ above the $B\bar{B}$ production threshold. It can be expected that a peak at this energy will also be observed in inelastic channels associated with annihilation into light hadrons and in transitions between different states of botomonium, see~\cite{Belle-II:2025ubm}.

The $\Upsilon(4S)$ resonance in the $B\bar{B}$ production cross section in our approach arises from the presence of a virtual level in the system of interacting $B$-mesons in our multichannel problem. As already mentioned in
Sec.~\ref{sec:DD} in relation to $D$-mesons, this result
does not contradict the generally accepted idea that $\Upsilon(4S)$
represents a bound state of $b\bar{b}$. The exact wave
function of  $\Upsilon(4S)$ meson contains not only the contribution of the $b\bar{b}$ state but also other contributions. For example, if we use the Cornell potential as a model of the interaction between the $b$ and $\bar{b}$ quarks, it turns out that the size of the $\Upsilon(4S)$ resonance is on the order of $1\,\mbox{fm}$. At such distances, the admixture of light quarks and gluons should already be noticeable. When calculating the matrix elements of different transitions, the dominant contribution will come from different
components of the total $\Upsilon(4S)$ wave function. In particular, the dominant contribution to the matrix element of the transition to the $B\bar{B}$ final state comes from the component of the $\Upsilon(4S)$ wave function containing
an admixture of light quarks. The interaction in this state is considered within the framework of our approach by introducing an effective potential for the interaction between $B$ mesons.

Our model also allows us to obtain predictions for the exclusive cross sections for the production of pairs of neutral and charged $B^{(*)}$ mesons,
and therefore for their ratios
\begin{equation}
R_{21}=\frac{\sigma(e^{+}e^{-}\to B^{0}\bar{B}^{0})}{\sigma(e^{+}e^{-}\to B^{+}B^{-})}\,,\, R_{43}=\frac{\sigma(e^{+}e^{-}\to B^{*0}\bar{B}^{0})}{\sigma(e^{+}e^{-}\to B^{+}B^{*-})}\,,\, R_{65}=\frac{\sigma(e^{+}e^{-}\to B^{*0}\bar{B}^{*0})}{\sigma(e^{+}e^{-}\to B^{*+}B^{*-})}\,.
\end{equation}

\begin{figure}[!tb]
	\centering
	\includegraphics[totalheight=5.1cm]{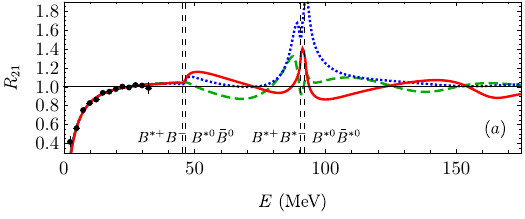}
	
	\includegraphics[totalheight=5.1cm]{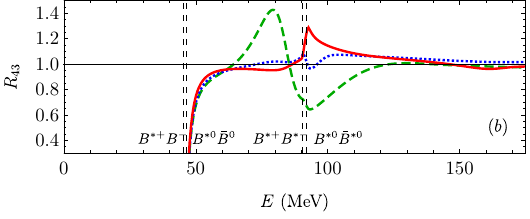}
	
	\includegraphics[totalheight=5.1cm]{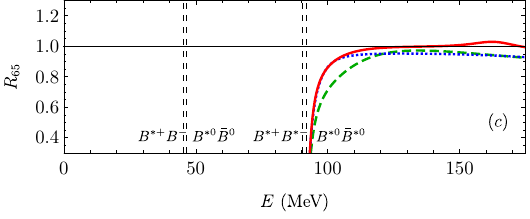}
	
	\caption{Predictions for the energy dependence of the ratios $R_{21}$~(a), $R_{43}$~(b), and $R_{65}$~(c) for three different sets of model parameters (see~\cite{Salnikov2026Charge}). The experimental points
		for $R_{21}$ were recalculated from the data in Ref.~\cite{Abumusabh2026Measurements}.
		The vertical dashed lines indicate the thresholds for the production of the corresponding
		states.}\label{fig:BBratios}
\end{figure}

Fig.~\ref{fig:BBratios} shows the energy dependence of these ratios for three different sets of model parameters. All three sets provide a good description of the available experimental data, but the predictions for the ratios $R_{ij}$ in the energy region where experimental data are not yet available differ significantly. It is clear that the ratios $R_{21}$ and $R_{43}$ can differ significantly from unity, especially near the $B^{*}\bar{B}^{*}$ production threshold. Detailed measurements of the effects of isotopic invariance violation in the processes $e^{+}e^{-}\to B^{(*)}\bar{B}^{(*)}$ will help clarify the structure of the $B^{(*)}$ meson interaction.

The different energy dependence of the cross sections for the production of charged and neutral $B^{(*)}$ mesons in $e^{+}e^{-}$ annihilation, associated with the interaction
in the final state, are important to consider when measuring the mass difference between these mesons.
In Refs.~\cite{Marciano1990, Lepage1990, Eichten1990, Kaizer2003, Voloshin2003, Voloshin2005}, the possible influence of isotopic invariance violation on the energy dependence of the $B^+B^-$ and $B^0\bar {B}^0$ production cross sections was studied in various ways. In our work, we showed that the systematic error in measuring the mass difference between the $B^{+}$ and $B^{0}$ mesons can reach $\delta M\sim 0{.}4\,{\text{MeV}}$ if measurements are made at the peak of the $\Upsilon(4S)$ resonance without properly accounting for the effects of isotopic invariance violation. Note that this value
$\delta M$ exceeds the stated uncertainty of the most accurate measurement to date
of the difference in the masses of $B^{+}$ and $B^{0}$ mesons, performed
at the BaBar detector~\cite{Aubert2008d} ($M_{B^{0}}-M_{B^{+}}=0{.}33\pm0{.}05\pm0{.}03\,{\text{MeV}}$).
In the BaBar measurement, only the difference in the phase volume of the final states for charged and neutral $B$ mesons was taken into account. Recent work by the Belle and Belle II collaborations \cite{Abumusabh2026Measurements} used direct measurements of the energy dependence of the ratio $R_{21}$  yielding the result $M_{B^{0}} - M_{B^{+}} =  0.495\pm0.024\pm0.005\,\text{MeV}$. This confirms the need to take into account the final-state interaction between $B$ mesons when measuring their masses.
 
\section{Conclusion}

Near-threshold resonances have been detected in the cross sections of a wide variety of processes involving the production of hadron pairs. The interaction of slow hadrons in the final state is a natural explanation of such features.
Indeed, in the vast majority of cases, taking into account the interaction
between hadrons allows us to successfully explain the experimentally observed
features in the energy dependence of the cross sections of various processes.
This makes it unnecessary to assume the existence of any new,
unknown particles produced in the intermediate state.

Resonances observed in processes involving the production of hadron pairs near the threshold, as in slow particle scattering processes, are insensitive to the fine details of the interaction between hadrons. This allows for various approaches to describing the final-state interaction and the use of different parametrization of the interaction potentials. The approach discussed here is based on solving the Schr\
\"odinger equation in the coordinate representation and finding the hadron pair wave functions near zero.
This approach allows for the easy inclusion of effects such as the Coulomb interaction, tensor forces, isotopic invariance violation, and transitions between several reaction channels with different thresholds.
Within this approach, effects associated with the interaction of a pair of virtual hadrons in an intermediate state can also be considered.

Various studies have shown that the nucleon-antinucleon interaction in the final state explains the strong energy dependence of the cross sections for the $e^{+}e^{-}\to p\bar{p}$ and $e^{+}e^{-}\to n\bar{n}$ processes near the thresholds, as well as the energy dependence of the ratios of the electromagnetic
form factors for the proton and neutron. Furthermore, taking into account the nucleon-antinucleon interaction in the intermediate state allows us to describe the sharp drop in the cross sections for the $e^{+}e^{-}\to 6\pi$ and $e^{+}e^{-}\to K^{+}K^{-}\pi^{+}\pi^{-}$ processes near the $N\bar{N}$ pair production threshold. The interaction between $D^{(*)}$
and $\bar{D}^{(*)}$ mesons (as well as $B^{(*)}$ and $\bar{B}^{(*)}$
mesons), taking into account several available reaction channels, explains the appearance of pronounced peaks in the cross sections of the $e^{+}e^{-}\to D^{(*)}\bar{D}^{(*)}$ processes (as well as $e^{+}e^{-}\to B^{(*)}\bar{B}^{(*)}$) and the deep dips
between these peaks. Studying these and other processes in the near-threshold
energy region allows us to obtain new information about the interaction
between hadrons at large distances.

\section*{Acknowledgement}
We are grateful to T.~Aushev, A.~Drutskoy, and G.~Shestakov for useful discussions.

\end{document}